\documentclass[a4paper,11pt]{article}
\pdfoutput=1 
\usepackage{jheppub} 
\usepackage{epsfig}
\usepackage{caption}
\usepackage{yfonts}
\usepackage{subcaption}
\usepackage{color,soul}
\usepackage{float}
\usepackage{amsfonts}
\usepackage{url}
\usepackage{slashed}
\usepackage{accents}
\usepackage{bbm,multicol}
\usepackage{lipsum}
\usepackage{mathtools}
\usepackage{physics}

\usepackage[normalem]{ulem}

\hypersetup{
  bookmarks=true,         % show bookmarks bar?
  unicode=false,          % non-Latin characters in Acrobat?s bookmarks
  pdftoolbar=true,        % show Acrobat?s toolbar?
 pdfmenubar=true,        % show Acrobat?s menu?
 pdffitwindow=true,     % window fit to page when opened
 pdfstartview={FitH},    % fits the width of the page to the window
 pdfsubject={Dark Matter},   % subject of the document
 pdfnewwindow=true,      % links in new window
 pdfcreator={RevTeX},
 colorlinks=true,       % false: boxed links; true: colored links
 linkcolor=red,          % color of internal links
 citecolor=blue,        % color of links to bibliography
 filecolor=black,      % color of file links
 urlcolor=blue,           % color of external links
  }

\title{Cosmology of complex scalar dark matter: interplay of self-scattering and annihilation}
\author[]{Avirup Ghosh,}
\author[]{Deep Ghosh}
\author[]{and Satyanarayan Mukhopadhyay}
\affiliation[]{School of Physical Sciences, Indian Association for the Cultivation of Science, 2A and 2B Raja S.C. Mullick Road, Kolkata 700 032}

\emailAdd{spsag2510@iacs.res.in}
\emailAdd{tpdg@iacs.res.in}
\emailAdd{tpsnm@iacs.res.in}

\abstract{The cosmology of a standard model (SM) gauge singlet complex scalar dark matter (DM), stabilized by a reflection symmetry, is studied including all renormalizable interactions that preserve the reflection symmetry but can break the larger global $U(1)$ symmetry of DM number. We find an interesting interplay of the ensuing DM self-scatterings and annihilations in generating the present DM density, and possible particle-antiparticle asymmetry in the DM sector. The role of DM self-scatterings in determining its present density and composition is a novel phenomenon. The simultaneous presence of the self-scatterings and annihilations is required to obtain a non-zero asymmetry, which otherwise vanishes due to unitarity sum rules.}

\begin{document} 
\maketitle
\flushbottom
\section{Introduction}
\label{sec:sec1}
Perhaps the simplest studied scenario for thermal dark matter (DM) is that of a real scalar field ($S$), which is a singlet under the standard model (SM) gauge interactions of $SU(3) \times SU(2) \times U(1)$~\cite{zee, McDonald}. The new degree of freedom added to the SM field content here is at an absolute minimum of one. Such a DM particle can be stable due to the existence of an effective reflection symmetry ($Z_2$ symmetry), under which the DM field is odd ($S \rightarrow -S$) and the SM particles are even. A renormalizable interaction of the real scalar singlet with the SM Higgs doublet ($H$) exists, namely, $S^2 |H|^2$, which is allowed by the reflection symmetry~\cite{zee, McDonald, Burgess, Patt}. The cosmology of this scenario is also very simple -- depending upon the mass of the singlet $S$, it may go through pair-annihilations to different SM particles, achieve and stay in thermodynamic equilibrium with the SM sector in the early Universe, and eventually freeze-out with a thermal relic abundance, which is determined by the two unknown parameters of the singlet mass and coupling to the Higgs. Such a scenario also presents with different signals at direct, indirect and collider searches -- and is strongly constrained by them in a large region of the parameter space providing the required DM abundance~\cite{cline, Athron:2017kgt, Athron:2018ipf}. 

In general, the DM particle may or may not be self-conjugate and it may be charged under a global symmetry of DM number. What new phenomena do we expect in such a case? Here, there are two broad possibilities. The first is that of a restricted scenario in which we impose the strict conservation of this global DM number symmetry, while the second possibility is of a more general scenario in which such a symmetry can be broken by different interaction terms. In the first case, the phenomena are essentially same as that of the self-conjugate DM, albeit with twice the new degrees of freedom. However, there is no strong argument behind the conservation of such global continuous symmetries. We know, for example, that it is very likely that the global $U(1)_{\rm B}$ symmetry of baryon number must be broken in order to dynamically generate the baryon asymmetry of the Universe~\cite{Weinberg:1979sa}. Similarly, the SM gauge quantum numbers of neutrinos allow for a Majorana mass term -- whose existence is being searched for -- and such a term would violate the global $U(1)_{\rm L}$ symmetry of lepton number. Such global symmetry breaking effects are generically found in unified theories of the strong, weak and electromagnetic interactions~\cite{Pati:1974yy, Georgi:1974sy}. It is also generally argued that global continuous symmetries may not be conserved  due to quantum gravitational effects~\cite{Witten:2017hdv}. Discrete symmetries, on the other hand, may remain in effective low energy theories, in some cases as a remnant of larger continuous symmetries~\cite{Witten:2017hdv}. 

Thus the second general possibility of including interactions that do not conserve the global symmetry of DM number should be explored in detail. And we expect to encounter new phenomena in studying the cosmology of such a scenario. A complex scalar singlet DM constitutes the version of the most minimal thermal DM scenario, with now a non-self-conjugate DM particle instead of a self-conjugate one. In this paper, we carry out a detailed study of this scenario, and find several interesting effects in its cosmological history, some of them novel. No complete study of such a general scenario for a complex scalar DM has been carried out so far, to our knowledge, and the previous studies of the complex scalar DM are restricted to symmetric weakly interacting massive particle (WIMP)-like behaviour~\cite{Barger:2008jx, Chiang:2017nmu}.  The different interaction terms present in such a scenario and their role in the various scattering processes are discussed in Sec.~\ref{sec:sec2} in the following.

The primary effect observed is that of the close interplay between the self-scattering and annihilation processes in determining the present DM density and composition. In particular, the fact that DM number may now be violated with all renormalizable interactions included, leads to the possibility of creation of a particle-antiparticle asymmetry in the DM sector. Such an asymmetry generation of course crucially depends on the CP-violation in the relevant scattering rates, which is non-zero only when all of the self-scattering and annihilation processes are simultaneously present. The absence of any one of them leads to the vanishing of the CP-violation, and hence the asymmetry -- by very general arguments of unitarity sum rules. This close interplay of the self-scatterings and annihilations is a very novel phenomenon, which we discuss in Sec.~\ref{sec:sec3}, after setting up the Boltzmann kinetic equations for the DM and anti-DM number densities. We then present the approximate analytic solutions to these equations (Sec.~\ref{sec:sec31}), which clearly demonstrate the role of different type of scattering processes in determining the net DM yields as well as any possible asymmetries, which are then compared with the general numerical solutions (Sec.~\ref{sec:sec32}). 

In Sec.~\ref{sec:sec4} we compute the CP-violation (at one-loop level) and the thermally averaged symmetric and asymmetric scattering rates in the complex singlet scenario, that are inputs to the kinetic equations, to understand the typical values of these quantities that one might expect in this model. We then go on to discuss the close interplay of the coupling parameters that determine the self-scattering and annihilation rates in fixing the DM relic density and asymmetry. We summarize our results in Sec.~\ref{sec:sec6}. The details of the CP-violation calculation for the sub-leading coversion process $\chi+\phi \rightarrow \chi^{\dagger}+\phi$ are provided in Appendix~\ref{App.A}.

%%%%%%%%%%%%%%%%%%%%%%%%%%%%%%%%%%%%%%%
\section{Gauge singlet complex scalar dark matter: general scenario}
\label{sec:sec2}
The most general low-energy effective Lagrangian of a SM gauge singlet complex scalar ($\chi$), that is odd under a $Z_2$ reflection symmetry (with $\chi \rightarrow -\chi$), is given as follows:
\begin{equation}
\mathcal{L} \supset (\partial_\mu \chi)^\dagger \partial^\mu \chi - m_{\chi}^2 \chi^\dagger \chi - \frac{1}{2} \left(\tilde{\mu}^2 \chi^2 + {\rm h.c.}\right)-\lambda_{\chi H} |\chi|^2 |H|^2,
\label{eq:lag0}
\end{equation}
where $H$ is the SM Higgs doublet. The interaction with the Higgs boson, however, is severely constrained by spin-independent direct detection probes for a large range of the DM mass, as well as by the search for invisible Higgs decay for a DM that is lighter than half of the Higgs mass. With such constrained values of the Higgs portal coupling, it is difficult to obtain the observed DM abundance in a large range of the DM mass. Therefore, we shall discuss the DM cosmology, assuming, for all practical purposes, that  the coupling $\lambda_{\chi H} \simeq 0$. As we shall see in the following, the effects of the Higgs coupling-like term in the DM cosmology will be captured by a new interaction that is then necessary for DM thermalization. The $\tilde{\mu}^2 \chi^2 $ term and its hermitian conjugate both break the larger global $U(1)_{\chi}$ symmetry of DM number under which the DM particle (denoted by $\chi$) has charge $+1$ and its antiparticle (denoted by $\chi^\dagger$) has a charge of $-1$.

We shall assume the $U(1)_{\chi}$ breaking $\tilde{\mu}^2$ term to be much smaller than the corresponding $U(1)_{\chi}$ conserving mass term $m_{\chi}^2$. In such a case, the DM mass eigenstates will essentially correspond to the particle states with $U(1)_{\chi}$ charge of $+1$ and $-1$, throughout the cosmological evolution of interest in this study. In late epochs, after the chemical decoupling of the DM species, particle-antiparticle oscillations can be caused by the presence of the $U(1)_{\chi}$ breaking mass term, whose effects are discussed in Refs.~\cite{Buckley:2011ye, Cirelli:2011ac, Tulin:2012re}. In particular, the oscillation probability is obtained to be $P(\chi\rightarrow\chi^{\dagger})= \sin^2\left({\mid\tilde{\mu}\mid^2 t}/{m_\chi}\right)$. In order for the oscillation dynamics to not affect the freeze-out process, the characteristic time scale of oscillations ($\tau=\left({\mid\tilde{\mu}\mid^2}/{m_\chi}\right)^{-1}$) should be larger than the freeze-out time scale ($t_F \sim H^{-1} (T=T_F)$), where $H(T=T_F)$ is the Hubble parameter at the freeze-out temperature. 
With $T_F \sim m_\chi$, requiring $\tau > t_F$ implies that as long as $|\tilde{\mu}|^2 \lesssim m_\chi^3/M_{\rm Pl}$, the mass and charge eigenstates can be taken to be essentially the same during the cosmological evolution time scale of the DM species. 

For example, with a typical order of the DM mass of $m_\chi = 10\, \rm TeV$, which is found to be relevant in the subsequent analyses, this upper bound on the possible size of $|\tilde{\mu}|$ is around $80$ keV. The upper bound on $|\tilde{\mu}|$ is modified appreciably in the presence of additional scattering processes which delay the onset of oscillations (see Ref.\cite{Cirelli:2011ac} for details). In our model, we find that for $|\tilde{\mu}|=10\, \rm GeV$ and $|\tilde{\mu}|=100 \,\rm GeV$, the oscillation starts at temperatures $T \sim1\, \rm GeV$ and $T \sim 10 \,\rm GeV$, respectively, for $m_\chi=10\,\rm TeV$, whereas the corresponding freeze-out dynamics takes place at  temperature $T \sim \mathcal{O}(500 \,\rm GeV)$. Therefore, we can safely carry out the subsequent analysis of the Boltzmann system in the charge basis, as the oscillation phenomenon is not relevant before or during the freeze-out process, as long as $|\tilde{\mu}|$ is around two orders of magnitude smaller than the DM mass.

With the condition that $\lambda_{\chi H} \simeq 0$, as discussed above, in order to thermalize the DM state with the SM sector, so that a viable thermal production mechanism for the DM density is obtained, we are forced to introduce a mediator degree of freedom in the scenario. A minimal new addition in such a case would be a real scalar field $\phi$, which is also a singlet under the SM gauge interactions, but is even under the $Z_2$ reflection symmetry. Since $\phi$ can couple to both the DM state as well as the SM sector, it can effectively bring the two sectors to thermal equilibrium. The additional interaction terms in the Lagrangian density are now as follows:
\begin{align}
-\mathcal{L_{\rm int}}  \supset  \mu  \chi^\dagger \chi  \phi+ \left(\frac{\mu_1}{2} \chi^2 \phi+ {\rm h.c.} \right)  + \frac{\lambda_1}{4} \left(\chi^\dagger \chi \right)^2 +   \left(\frac{\lambda_2}{4!}\chi^4 +{\rm h.c.} \right)+\left(\frac{\lambda_3}{4}\chi^2 \phi^2 +{\rm h.c.} \right)\nonumber\\
+\left(\frac{\lambda_4}{3!}\chi^3\chi^{\dagger} +{\rm h.c.} \right)
+ \frac{\lambda_5}{2} \phi^2 \chi^\dagger \chi  +  \frac{\mu_\phi}{3!} \phi^3 +  \frac{\lambda_\phi}{4!} \phi^4 + \frac{\lambda_{\phi H}}{2} \phi^2 |H|^2 + \mu_{\phi H} \phi |H|^2\,\,\,
\label{eq:lag}
\end{align}
In this most general renormalizable interaction Lagrangian involving the $\chi$ and $\phi$ fields, the interaction terms with couplings $\mu_1, \lambda_2, \lambda_3$ and $\lambda_4$ break the global $U(1)_{\chi}$ symmetry, in addition to the $\tilde{\mu}^2$ term in Eq.~\ref{eq:lag0} above. All of these terms, in general, can also have complex coupling parameters, one of which can be chosen to be real by an appropriate redefinition of the $\chi$ field. 

The scalar mass and quartic terms are chosen appropriately to ensure that the $\chi$ and $\phi$ fields do not obtain any non-zero vacuum expectation values, where the former condition also implies that the $Z_2$ reflection symmetry remains unbroken, rendering the DM stable. With this condition, the couplings between the $\phi$ and the SM Higgs fields will eventually lead to $\phi$ mixing with the Higgs particle after electroweak symmetry breaking through the $\phi |H|^2$ term only. Therefore, the current Higgs coupling measurements will strongly constrain the $\mu_{\phi H}$ coupling. On the other hand, the $\lambda_{\phi H}$ coupling is not as strongly bounded, even if $\phi$ is much lighter than the Higgs. This is because, if the Higgs boson decays to two $\phi$ particles, each of the resulting $\phi$ will further decay to visible SM final states through its small mixing with the Higgs. If the decay width of $\phi$ is so small that the $\phi$ particles decay outside the LHC detectors, there will be an upper bound on $\lambda_{\phi H}$ from the search for invisible decays of the Higgs boson. Using the $95 \%$ C.L. upper limit on the Higgs invisible branching ratio obtained by a combination of the $7,8$ and $13$ TeV data from the CMS collaboration~\cite{CMS:2018yfx}, we find $\lambda_{\phi H} \lesssim 10^{-2}$. Such couplings are however sufficient for the $ \lambda_{\phi H} $ and $\mu_{\phi H}$ terms to keep the $\phi$ particles in thermal equilibrium with the SM sector. 

The small $\phi$ mixing with the Higgs however can induce signals in DM direct detection probes through the $\chi^\dagger \chi  \phi$ and $\chi^2 \phi$ terms, which will induce $\chi$ scattering with nucleons. Therefore, to a first approximation, we also set $\mu \simeq 0$ and $\mu_1 \simeq 0$. As we shall see in the following, the essential aspects of the cosmology of $\chi$ particles are effectively captured even in such a case. With these parameter choices, we have checked that the $U(1)_\chi$ symmetry breaking interaction terms do not generate a large $\tilde{\mu}^2$ term through loop corrections.\footnote{As is well-known from the example of scalar field theory with a $\phi^4$ interaction, at one-loop level, the quartic couplings do not generate any quantum corrections to the $\tilde{\mu}^2$ term. Furthermore, since we have also set the trilinear couplings to be vanishing to begin with, the leading quantum correction to the $\tilde{\mu}^2$ term will be generated only at the two loop level. Such two-loop corrections to $|\tilde{\mu}|$ are, however, expected to be at least two orders of magnitude smaller than $m_\chi$, which is the typical upper bound required for oscillation phenomenon to occur only much after the DM freeze-out.}. We can thus consistently treat the $U(1)_\chi$ charge eigenstates as the DM mass eigenstates for the cosmological evolution until the freeze-out time scale, which is the subject matter of this study. For the astrophysical dynamics in later epochs, the effect of these terms should be included.

Through their interactions with $\phi$, the DM particles and anti-particles are assumed to remain in kinetic equilibrium with the SM thermal bath throughout the evolution of the chemical processes that determine its density. The chemical reactions that affect the DM density and composition, following from the interactions in Eq.~\ref{eq:lag} are as follows:
\begin{enumerate}
\item {\em Self-scattering reaction leading to particle-antiparticle conversion}: $\chi+\chi \rightarrow \chi^{\dagger} + \chi^{\dagger}$. Due to the presence of a complex coupling parameter, this reaction can be CP-violating and can create  a particle-antiparticle asymmetry in the DM sector. Moreover, it violates the DM number of $U(1)_{\chi}$ by four units, but cannot change the total DM and anti-DM number. 

\item {\em Self-scattering reaction leading to particle-antiparticle conversion}: $\chi+\chi \rightarrow \chi + \chi^{\dagger}$. Similar to the previous process, this can also lead to CP-violation and asymmetry in the DM sector. This process violates the DM number by two units, but does not modify the total DM number.

\item {\em DM number violating annihilation}: $\chi+\chi \rightarrow \phi+\phi$. This process can also violate CP-symmetry and create asymmetric DM, due to the complex coupling of $\lambda_3$. It violates DM number by two units, and also changes the total DM and anti-DM number by the same amount.

\item {\em DM number conserving annihilation}: $\chi+\chi^\dagger \rightarrow  \phi+\phi$. This process is CP-conserving, and cannot by itself create a DM-anti-DM asymmetry. However, as shown in Ref.~\cite{Ghosh:2021ivn}, it can indirectly affect the generated asymmetry by controlling the out-of-equilibrium number density of relevant particles in the thermal bath. This process conserves the $U(1)_{\chi}$ DM number symmetry, but changes the total DM and anti-DM number by two units. 
\end{enumerate}

In addition to the processes included in the DM and anti-DM number-density evolution equations, the process, $\chi+\phi \rightarrow \chi^{\dagger}+\phi$ can also change the DM number by $-2$ units, whereas its CP-conjugate process,  $\chi^{\dagger}+\phi\rightarrow \chi+\phi$ changes the DM number by $+2$ units. Therefore, if the CP-violation in this process is substantial, it can lead to particle-antiparticle conversions, thereby affecting the DM asymmetry. However, as discussed in Sec.~\ref{sec:sec4}, and detailed in Appendix~\ref{App.A}, we find the contribution of this process to the CP-violation to be negligible in the parameter region of interest in this study, and therefore, we have not included the effect of this scattering process in generating the DM asymmetry.

%%%%%%%%%%%%%%%%%%%%%%%%%%%%%%%%%%%%%%%
\section{Boltzmann equations and unitarity sum rules}
\label{sec:sec3}
We shall compute, in Sec.~\ref{sec:sec4}, the reaction rates for the  processes described in the previous section, and the possible CP-violation in them in terms of the coupling and mass parameters appearing in Eqs.~\ref{eq:lag0} and \ref{eq:lag}. However, to understand the evolution dynamics  of the number densities of the $\chi$ and $\chi^\dagger$ particles in the thermal bath, it is useful to first study the kinetic equations with the thermally averaged reaction rates as inputs. Through this process, we shall arrive at an understanding of the relationships between these reaction rates and their consequences --- in the DM density, the CP-violation in the above reactions, and the resulting possible particle-antiparticle asymmetry in the DM sector. We take up this study in this section.

The Boltzmann kinetic equations~\cite{Kolb:1990vq} governing the number densities are given as follows:
\begin{align}
\frac{dn_\chi}{dt}+3Hn_\chi = -\int \prod^{4}_{i=1} \frac{d^3 p_i}{(2\pi)^3 2 E_{p_i}}g^2_{\chi}   (2 \pi)^4 \delta^{(4)}(p_1+p_2-p_3-p_4)\bigg[2\bigg(|M|^2_{\chi\chi \rightarrow \chi^{\dagger}\chi^{\dagger}}f_\chi (p_1)f_\chi(p_2)\nonumber\\
-|M|^2_{\chi^{\dagger}\chi^{\dagger} \rightarrow \chi\chi}f_{\chi^{\dagger}} (p_3)f_{\chi^{\dagger}}(p_4)\bigg)
+\bigg(|M|^2_{\chi\chi \rightarrow \chi^{\dagger}\chi}f_\chi (p_1)f_\chi(p_2)-|M|^2_{\chi^{\dagger}\chi \rightarrow \chi\chi}f_{\chi^{\dagger}} (p_3)f_\chi(p_4)\bigg)\nonumber\\
-\bigg(|M|^2_{\chi^{\dagger}\chi^{\dagger} \rightarrow \chi^{\dagger}\chi}f_{\chi^{\dagger}} (p_1)f_{\chi^{\dagger}}(p_2)-|M|^2_{\chi^{\dagger}\chi \rightarrow \chi^{\dagger}\chi^{\dagger}}f_{\chi^{\dagger}} (p_3)f_\chi(p_4)\bigg)\nonumber\\
+2\bigg(|M|^2_{\chi\chi \rightarrow \phi\phi}f_\chi (p_1)f_\chi(p_2)-|M|^2_{\phi\phi \rightarrow \chi\chi}f_\phi (p_3)f_\phi(p_4)\bigg)\nonumber\\
+\bigg(|M|^2_{\chi^{\dagger}\chi \rightarrow \phi\phi}f_{\chi^{\dagger}} (p_1)f_\chi(p_2)-|M|^2_{\phi\phi \rightarrow \chi^{\dagger}\chi}f_{\phi} (p_3)f_\phi(p_4)\bigg)\bigg].
\label{eq:boltz}
\end{align} 
Here, $H$ denotes the Hubble expansion parameter, $f_i(p_i)$ are the distribution functions of the particle type $i$, and $g_\chi$ is the number of internal degrees of freedom of the $\chi$ field in general, which is equal to $1$ in the present case. The corresponding equation for $n_{\chi^\dagger}$ can be obtained by replacing $\chi$ by $\chi^\dagger$ everywhere in the above equation, and vice versa. We have not explicitly shown the appropriate symmetry factors for identical particles in the initial and final states, which have, however, been incorporated in our computations in the following. 

In terms of the standard dimensionless variables $Y_\chi=n_\chi/s$ and $x=m_\chi/T$, where $s$ and $T$ are the entropy density and temperature of the radiation bath, respectively, the Boltzmann equations can be re-written as
\begin{align}
\frac{dY_\chi}{dx} &= -\frac{s}{2 H x}\bigg[\left(\expval{\sigma v}_1+\frac{\expval{\sigma v}_2}{2}+\expval{\sigma v}_3\right)\left(Y^2_\chi-Y^2_{\chi^{\dagger}}\right)+\expval{\sigma v}_3\left(Y^2_{\chi^{\dagger}}-Y^2_0\right)\nonumber\\
&+ \expval{\epsilon\sigma v}_1\left(Y^2_{\chi^{\dagger}}-Y^2_0\right)+\expval{\epsilon\sigma v}_2\left(Y_{\chi^{\dagger}}Y_{\chi}+\frac{Y^2_{\chi^{\dagger}}}{2}-\frac{Y^2_\chi}{2}-Y^2_0\right)+2\expval{\sigma v}_A \left(Y_{\chi^{\dagger}}Y_{\chi}-Y^2_0\right)\bigg]\nonumber\\
\frac{dY_{\chi^{\dagger}}}{dx} &= -\frac{s}{2 H x}\bigg[\left(\expval{\sigma v}_1+\frac{\expval{\sigma v}_2}{2}+\expval{\sigma v}_3\right)\left(Y^2_{\chi^{\dagger}}-Y^2_{\chi}\right)+\expval{\sigma v}_3\left(Y^2_{\chi}-Y^2_0\right)\nonumber\\
&- \expval{\epsilon\sigma v}_1\left(Y^2_{\chi}-Y^2_0\right)-\expval{\epsilon\sigma v}_2\left(Y_{\chi^{\dagger}}Y_{\chi}+\frac{Y^2_\chi}{2}-\frac{Y^2_{\chi^{\dagger}}}{2}-Y^2_0\right)+2\expval{\sigma v}_A \left(Y_{\chi^{\dagger}}Y_{\chi}-Y^2_0\right)\bigg].\nonumber\\
\label{eq:boltz_p}
\end{align}
Here,  $Y_0 =  \frac{1}{s} \int  \frac{d^3 p} {(2\pi)^3 } g_\chi f_0(p)$, with $f_0(p)=e^{-\frac{E(p)}{T}}$ being the equilibrium distribution function with zero chemical potential. We have also introduced the thermally averaged symmetric and asymmetric reaction rates $\expval{\sigma v}_f$ and $\expval{\epsilon \sigma v}_f$ for a particular final state $f$, respectively, where the latter is defined as follows:
\begin{equation}
\expval{\epsilon \sigma v}_{f} = \dfrac{\int \prod^{4}_{i=1} \frac{d^3 p_i}{(2\pi)^3 2 E_{p_i}}  (2 \pi)^4 \delta^{(4)}(p_1+p_2-p_3-p_4) \,\epsilon_f(p_i) |M_0|^2_f f_0(p_1)f_0(p_2)}{\int \dfrac{d^3 p_1}{(2\pi)^3} \dfrac{d^3 p_2}{(2\pi)^3} f_0(p_1)f_0(p_2)} \hspace{0.5cm},
\label{eq:cross}
\end{equation}
with   $|M_0|^2_f = |M|^2_{\chi\chi\rightarrow f}+|M|^2_{\chi^{\dagger}\chi^{\dagger}\rightarrow f^{\dagger}} $, and $\epsilon_f(p_i)$ is given by
\begin{align}
\epsilon_f (p_i) =\frac{ |M|^2_{\chi\chi\rightarrow f}-|M|^2_{\chi^{\dagger}\chi^{\dagger}\rightarrow f^{\dagger}}}{|M|^2_{\chi\chi\rightarrow f}+|M|^2_{\chi^{\dagger}\chi^{\dagger}\rightarrow f^{\dagger}}}.
\label{eq:def_eps}
\end{align} 
The corresponding thermally averaged symmetric reaction rates $\expval{\sigma v}_f$ can be computed using Eq.~\ref{eq:cross} by removing the $\epsilon_f(p_i)$ factors. Finally, we denote the reaction rates for $\chi+\chi \rightarrow \chi^{\dagger} + \chi^{\dagger}$ by $\expval{\sigma v}_1$ and  $\expval{\epsilon \sigma v}_1$, for $\chi+\chi \rightarrow \chi + \chi^{\dagger}$ by $\expval{\sigma v}_2$ and $\expval{\epsilon \sigma v}_2$ and for $\chi+\chi \rightarrow \phi+\phi$  by $\expval{\sigma v}_3$ and $\expval{\epsilon \sigma v}_3$. For the CP-conserving reaction $\chi+\chi^\dagger \rightarrow  \phi+\phi$ we denote the average reaction rate by $\expval{\sigma v}_A$.

We have parametrized the CP-violation resulting from the first three reactions discussed in Sec.~\ref{sec:sec2} in terms of $\expval{\epsilon \sigma v}_1$ and $\expval{\epsilon \sigma v}_2$ in Eqs.~\ref{eq:boltz_p}, and have eliminated $\expval{\epsilon \sigma v}_3$ from the Boltzmann equations using unitarity sum rules. This stems from the fact that the amplitudes for the CP-violating processes are related by CPT and S-matrix unitarity as follows~\cite{Kolb:1979qa, Baldes:2014gca, Baldes:2015lka}:
\begin{align}
\sum_f \int dPS_f |M|^2_{\chi\chi \rightarrow f }=\sum_f \int dPS_f  |M|^2_{f \rightarrow \chi\chi}= \sum_f \int dPS_f  |M|^2_{\chi^{\dagger}\chi^{\dagger}\rightarrow f^{\dagger}}, 
\label{eq:sum}
\end{align}
where we have used the S-matrix unitarity in the first equality, and CPT conservation in the second one. Here, the integral over $dPS_f$ sum over the momenta of the particles in the state $f$, as well as any discrete label that may be carried by $f$. There is a further sum over all possible final states $f$ that may be obtained starting from the initial state of $\chi \chi$, which has been explicitly indicated. This unitarity sum rule in Eq.~\ref{eq:sum}, together with the definition of the CP-violation parameter $\epsilon_f$ in Eq.~\ref{eq:def_eps}, imply the following sum rule relating the CP-violation in all the channels

\begin{align}
\sum_f \int dPS_f\,\, \epsilon_f\, |M_0|^2_f = 0, 
\label{eq:eps_unitarity}
\end{align}
where, the quantity $|M_0|^2_f $ has been defined above. Using this sum rule in Eq.~\ref{eq:eps_unitarity} for the three relevant CP-violating reactions with the initial state of $\chi \chi$ and final states in the set $f=\{\chi^{\dagger} + \chi^{\dagger}, \chi + \chi^{\dagger}, \phi+\phi\}$, we can express one of the asymmetric rates $\expval{\epsilon \sigma v}_f$ in terms of the other two, and therefore, can write the Boltzmann equations in terms of only two asymmetric reaction rates as in Eq.~\ref{eq:boltz_p}.

%%%%%%%%%%%%%%%%%%%%%%%%%%%%%%%%%%%%%%%
\subsection{Analytic solutions}
\label{sec:sec31}
We now discuss approximate analytic solutions to the Boltzmann equations~\ref{eq:boltz_p}. For this, it is convenient to define the symmetric and asymmetric yields, $Y_S = Y_\chi+Y_{\chi^\dagger}$ and $Y_{\Delta \chi}=Y_\chi-Y_{\chi^\dagger}$, respectively, and rewrite the Boltzmann equations in terms of these variables as
\begin{align}
\dfrac{dY_S}{dx}&=-\dfrac{s}{2Hx}\left[\expval{\sigma v}_A\left(Y^2_S-Y^2_{\Delta \chi}-4Y^2_0\right)+\expval{\sigma v}_3\bigg(\dfrac{Y^2_S+Y^2_{\Delta \chi}-4Y^2_0}{2}\bigg)-\expval{\epsilon\sigma v}_S Y_S Y_{\Delta \chi}\right]\nonumber\\
\dfrac{dY_{\Delta \chi}}{dx}&=-\dfrac{s}{2Hx}\left[\expval{\epsilon\sigma v}_S\left(\dfrac{Y^2_S-4Y^2_0}{2}\right)+\expval{\epsilon\sigma v}_D \dfrac{Y^2_{\Delta \chi}}{2}+\expval{\sigma v}_{all}\,Y_S Y_{\Delta \chi}\right].
\label{eq:asym}
\end{align} 
In order to write these equations in a compact form, we have further defined the following quantities: $\expval{\epsilon\sigma v}_S =\expval{\epsilon \sigma v}_1+\expval{\epsilon \sigma v}_2 $, $\expval{\epsilon\sigma v}_D=\expval{\epsilon \sigma v}_1-\expval{\epsilon \sigma v}_2$ and $\expval{\sigma v}_{all}=2\expval{\sigma v}_1+\expval{\sigma v}_2+\expval{\sigma v}_3$. Let us discuss the analytic solutions in two different regions of the scaled temperature variable $x$. In the first region, we consider values of $x$ upto its freeze-out value of $x_F=m_\chi/T_F$, i.e., $1\leq x \leq x_F$. Here, at the temperature $T_F$, all the relevant chemical reactions involving the DM particle freeze-out.  In the second region, we consider values $x>x_F$. 

For $1\leq x \leq x_F$, we can parametrize the number densities of the DM and anti-DM particles as 
\begin{equation}
Y_{\chi(\chi^\dagger)} =Y_0(1+\delta_{1(2)}),
\end{equation}
where, $\delta_{1}$ and $\delta_2$ parametrize the small deviations away from their common equilibrium value with zero chemical potential, $Y_0$. We expect that in this region, the deviations will satisfy the condition $|\delta_{1,2}|<< 1$. In the presence of CP-violating scatterings, in general, $|\delta_1| \neq |\delta_2|$. In terms of this parametrization, we have $Y_S = 2Y_0 (1+\delta)$ and 
$Y_{\Delta\chi}=Y_0 \bar{\delta}$, where $\delta=(\delta_1+\delta_2)/2$ and $\bar{\delta}=\delta_1-\delta_2$. 

Dropping terms quadratic in the small deviations, i.e., terms proportional to  $\delta^2$, $\bar{\delta}^2$ and $\delta \bar{\delta}$, we can solve Eqs.~\ref{eq:asym} algebraically. The total DM and anti-DM yield is obtained to be as follows:
\begin{equation}
Y_S(x) = 2Y_0\left[1+\dfrac{Hx}{ s Y_0}\dfrac{\expval{\sigma v}_{all}}{2\expval{\sigma v}_{ann}\expval{\sigma v}_{all}+\expval{\epsilon \sigma v}^2_S}\right]~~~~{\rm (for~} 1\leq x \leq x_F) ,
\label{eq:analytic1}
\end{equation}
whereas, the yield for the DM-anti-DM asymmetry is given by
\begin{equation}
|Y_{\Delta\chi}(x)|= \dfrac{2 H x}{s}\,\dfrac{\expval{\epsilon\sigma v}_S}{2\expval{\sigma v}_{ann}\expval{\sigma v}_{all}+\expval{\epsilon \sigma v}^2_S}~~~~{\rm (for~} 1\leq x \leq x_F),
\label{eq:analytic2}
\end{equation}
where, we have defined $\expval{\sigma v}_{ann} = \expval{\sigma v}_A + \expval{\sigma v}_3/2 $. In deriving these expressions, we have taken the non-relativistic form of $Y_0$, which is applicable near the freeze-out temperature. We can draw several important conclusions from these solutions:
\begin{enumerate}
\item The CP-conserving DM and anti-DM annihilation process, $\chi+\chi^\dagger \rightarrow  \phi+\phi$, can indirectly control the asymmetric yield. This can be seen in Eq.~\ref{eq:analytic2}, where we find that $|Y_{\Delta\chi}(x)|$ is inversely proportional to $\expval{\sigma v}_A $. This is because the out-of-equilibrium number density of the relevant species can be controlled by this CP-conserving reaction, which, in turn, controls the asymmetric yield. We have discussed the role of CP-conserving processes in producing cosmological particle-antiparticle asymmetries in detail in Ref.~\cite{Ghosh:2021ivn}.

\item In our parametrization of CP-violation in the Boltzmann equations~\ref{eq:boltz_p}, $\expval{\epsilon\sigma v}_S =\expval{\epsilon \sigma v}_1+\expval{\epsilon \sigma v}_2$ acts as a source term for the asymmetry, as seen in Eq.~\ref{eq:analytic2}. This is because, as explained earlier,  $\expval{\epsilon \sigma v}_3$ cannot be independently varied, and is determined in terms of $\expval{\epsilon \sigma v}_1$ and $\expval{\epsilon \sigma v}_2$ through the unitarity sum rules in Eq.~\ref{eq:eps_unitarity}. 

\item We see in Eq.~\ref{eq:analytic1}  that the the total DM yield is significantly affected by the pair annihilation process $\chi+\chi^\dagger \rightarrow  \phi+\phi$. This is because, we expect $\expval{\epsilon \sigma v}^2_S << \expval{\sigma v}_{ann}\expval{\sigma v}_{all}$, as the asymmetric scattering rates stem from the interference of tree and loop level amplitudes, and are therefore suppressed compared to the symmetric scattering rates. With this approximation, we have
\begin{equation}
Y_S \sim 2Y_0 \left(1+\dfrac{Hx}{sY_0 (2\expval{\sigma v}_A+ \expval{\sigma v}_3)} \right)
\label{eq:YS_approx_1}
\end{equation}
\end{enumerate}
In the second region with $x>x_F$, we can no longer ignore the terms quadratic in the deviations $\delta$ and $\bar{\delta}$, and the analytic solutions in closed form become cumbersome. However, the results are tractable under certain simplifying assumptions. We note that for $x > x_F$, the self-scattering reactions that can generate the asymmetry are mostly decoupled since the annihilation rate is proportional to the DM velocity, as detailed in the next section. Thus we see that in the Boltzmann equation for $Y_{\Delta \chi}$ in Eq.~\ref{eq:asym} the terms in the right hand side (RHS) proportional to $\expval{\epsilon\sigma v}_S$ and $\expval{\epsilon\sigma v}_D$ will be very small as well. Furthermore, since $Y_S \propto 1/\expval{\sigma v}_A$, the last term in the RHS will be proportional to the ratio $\expval{\sigma v}_{all}/\expval{\sigma v}_A$. Therefore, in the regime $\expval{\sigma v}_{all} << \expval{\sigma v}_A$, i.e., when the dominant scattering process is the CP-even annihilation, $Y_{\Delta \chi}$ essentially remains a constant, being frozen at its value around $x=x_F$. With this as input, we can solve the equation for $Y_S$ in Eq.~\ref{eq:asym}, and for dominantly s-wave contribution to $\expval{\sigma v}_A$, the solution is given by
\begin{align}
Y_S(x>x_F) &= |Y^F_{\Delta \chi}|\,\,\dfrac{1+r_F \exp[\lambda\, |Y^F_{\Delta \chi}|\,(x^{-1}-x^{-1}_F)]}{1-r_F \exp[\lambda\, |Y^F_{\Delta \chi}|\,(x^{-1}-x^{-1}_F)]},
\label{eq:YS_approx_2}
\end{align}
where,
\begin{equation}
r_F =\dfrac{Y^F_S-|Y^F_{\Delta \chi}|}{Y^F_S+|Y^F_{\Delta \chi}|}.
\end{equation}
Here, $\lambda = 1.32\, m_\chi M_{Pl}\,g^{1/2}_{*}\expval{\sigma v}_A$, and $Y^F_S$ and $Y^F_{\Delta \chi}$ represent the symmetric and asymmetric yields of DM obtained at $x=x_F$ using Eqs.~\ref{eq:analytic1} and~\ref{eq:analytic2}. 
We shall compare these approximate analytical results with the numerical solutions in the next sub-section.

%%%%%%%%%%%%%%%%%%%%%%%%%%%%%%%%%%%%%%%
\subsection{Numerical results}
\label{sec:sec32}
\begin{figure}[htb!]
\centering
\includegraphics[scale=0.5]{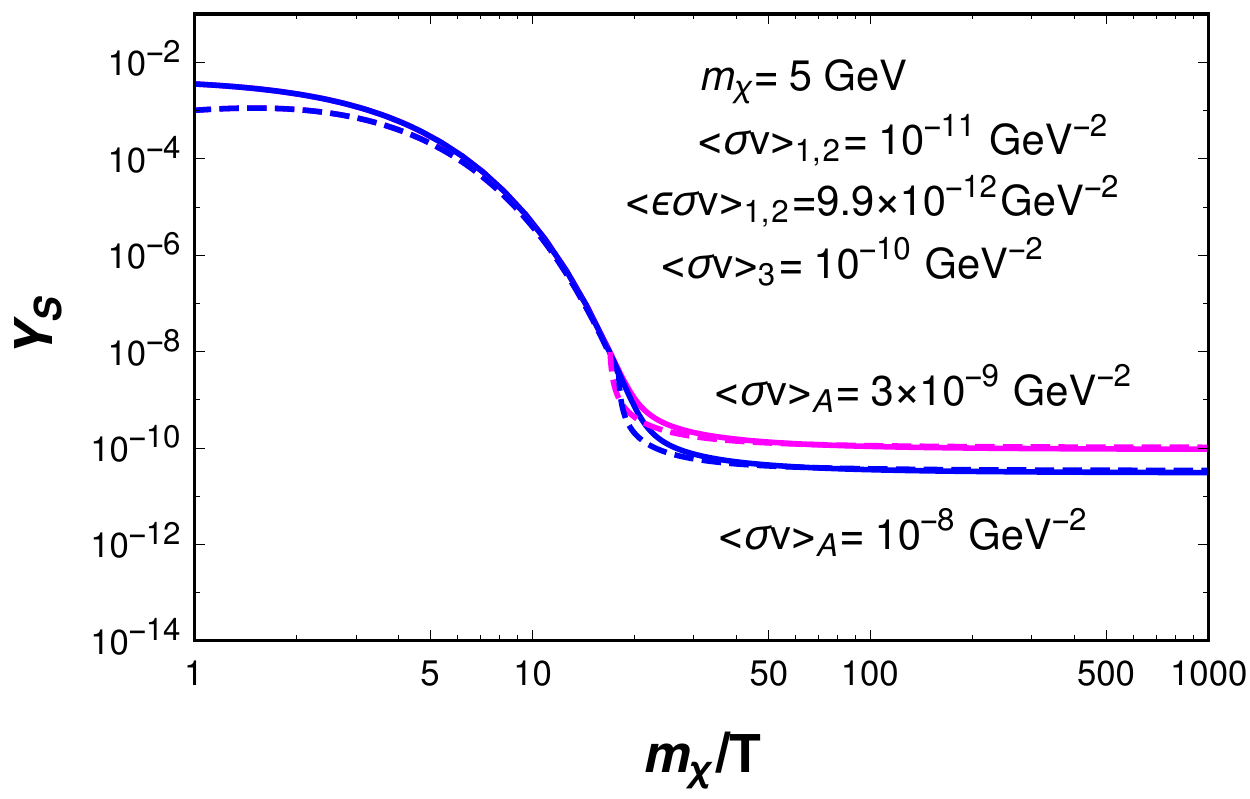}
\includegraphics[scale=0.5]{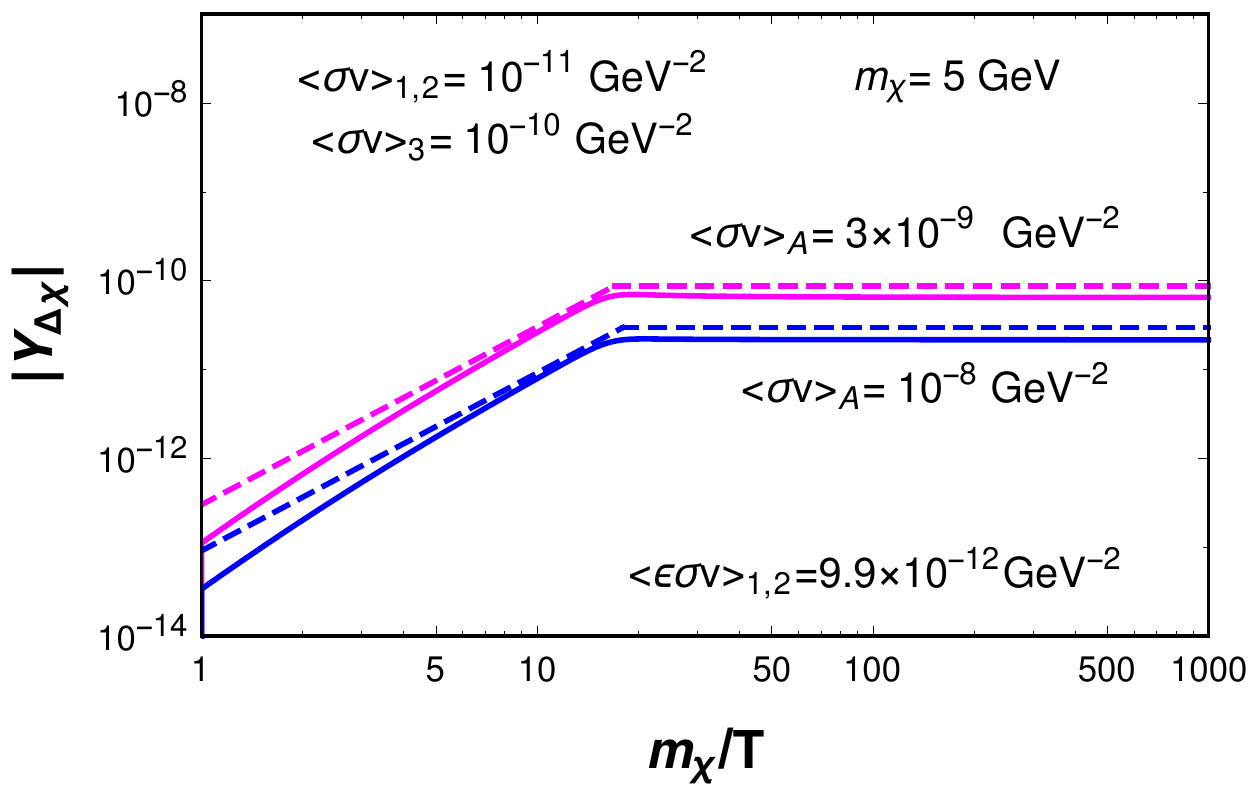}
\caption{\small{\em Evolution of the symmetric (left panel, with $Y_S=Y_\chi+Y_{\chi^\dagger}$) and the asymmetric yields (right panel, with $|Y_{\Delta \chi}|=|Y_\chi-Y_{\chi^\dagger}|$), as a function of the scaled inverse temperature variable $m_\chi/T$. The solid lines in both the figures represent the exact numerical solution of Eqs.~\ref{eq:boltz_p}. The dashed lines show the predictions based on the approximate analytic solutions, namely, Eqs.~\ref{eq:analytic1} and \ref{eq:analytic2}. The dark matter mass and CP-violating symmetric and asymmetric reaction rates have been kept fixed. We have shown the results for two different choices of the CP-conserving pair annihilation rate $\expval{\sigma v}_A $ for the process $\chi+\chi^{\dagger} \rightarrow \phi + \phi$ (by the blue and pink lines) to demonstrate its impact not only on the total yield, but also in indirectly controlling the particle anti-particle asymmetry as well.
}}
\label{fig:evolve}
\end{figure}

Having obtained the approximate analytic solutions to the sum and difference of DM and anti-DM yields for different ranges of the temperature in the thermal bath, we now go on to solve Eqs~\ref{eq:boltz_p} numerically to understand the exact relationship between the DM properties at freeze-out and the microscopic parameters. In Fig.~\ref{fig:evolve} we have shown the evolution of the symmetric (left panel, with $Y_S=Y_\chi+Y_{\chi^\dagger}$) and the asymmetric yields (right panel, with $|Y_{\Delta \chi}|=|Y_\chi-Y_{\chi^\dagger}|$), as a function of the scaled inverse temperature variable $m_\chi/T$. The solid lines in both the figures represent the exact numerical solution of Eqs.~\ref{eq:boltz_p}. The dashed lines show the predictions based on the approximate analytic solutions, namely, Eqs.~\ref{eq:analytic1} and \ref{eq:analytic2}. The dark matter mass and CP-violating symmetric and asymmetric reaction rates have been kept fixed. We have shown the results for two different choices of the CP-conserving pair annihilation rate $\expval{\sigma v}_A $ for the process $\chi+\chi^{\dagger} \rightarrow \phi + \phi$, by the blue and pink lines, to demonstrate its impact not only on the total yield, but also in the particle anti-particle asymmetry as well. As we can see, by increasing $\expval{\sigma v}_A $, one obtains a lower value of $|Y_{\Delta \chi}|$ throughout the evolution. We also see from Fig.~\ref{fig:evolve} (right panel) that a small out-of-equilibrium asymmetry $|Y_{\Delta \chi}|$ is generated from around $x \sim 1$, which eventually grows and freezes at around $x \sim x_F$. 

The analytic approximation is found to be in good agreement with the exact numerical solutions, the maximum difference between the two being of the order of $10\%$. The larger differences are found for smaller values of $x$. This is because in order to obtain an approximate closed form analytic solution to Eqs.~\ref{eq:boltz_p}, we have taken $dY_0/dx \sim -Y_0$, and have dropped an additional term proportional to $1/x$. This approximation is justified for $x > 1$, which is the main region of interest for determining the relic abundance. However, it is no longer a valid approximation for $x \sim 1$, where the additional $x-$dependent term also contributes, and an accurate (better than the $10\%$ accuracy obtained here) closed form solution cannot be obtained for $x \sim 1$ while retaining such a term. 

We note that the patching of the analytic solutions in the regions $1 \leq x \leq x_F$ and $x>x_F$ requires an input of the value of $x_F$ itself, and is therefore sensitive to how precisely we can estimate $x_F$. However, the asymptotic values of the yields $Y_S$ and $|Y_{\Delta \chi}|$ are found to be well approximated by the analytic solutions. Here, we have estimated $x_F$ as follows. Since we have $Y_S = 2Y_0 (1+\delta)$, and in the first patch the net deviation from equilibrium $\delta < 1$, we can use this condition in the expression for $Y_S$ in Eq.~\ref{eq:YS_approx_1}, and obtain a relation which determines the boundary of the two regions $x_F$ as
\begin{equation}
\frac{H(x_F) x_F}{s(x_F)Y_0(x_F) (2\expval{\sigma v}_A+\expval{\sigma v}_3)} \sim 1.
\end{equation} 
An iterative solution of this equation gives a good estimate of the value of the scaled freeze-out temperature $x_F$.

%%%%%%%%%%%%%%%%%%%%%%%%%%%%%%%%%%%%%%%
\section{CP-violation and scattering rates in the complex singlet model}
\label{sec:sec4}
We now compute the relevant CP-violation parameters and the thermally averaged symmetric and asymmetric scattering rates in the complex scalar singlet model, as a function of the model parameters. We then go on to study the dependence of the DM relic abundance and the possible particle-antiparticle asymmetry in the DM sector on such parameters, with a particular emphasis on the interplay between the DM self-scattering and annihilation in controlling the DM density and composition. 

We first  take up the three CP-violating processes one by one, and discuss the corresponding results at next-to-leading order in perturbation theory, following which we show the leading order results for the CP-conserving process.
\section*{Process 1: Self-scattering $\chi+\chi \rightarrow \chi^{\dagger} + \chi^{\dagger}$}

\begin{figure}[htb!]
\centering
\includegraphics[scale=0.39]{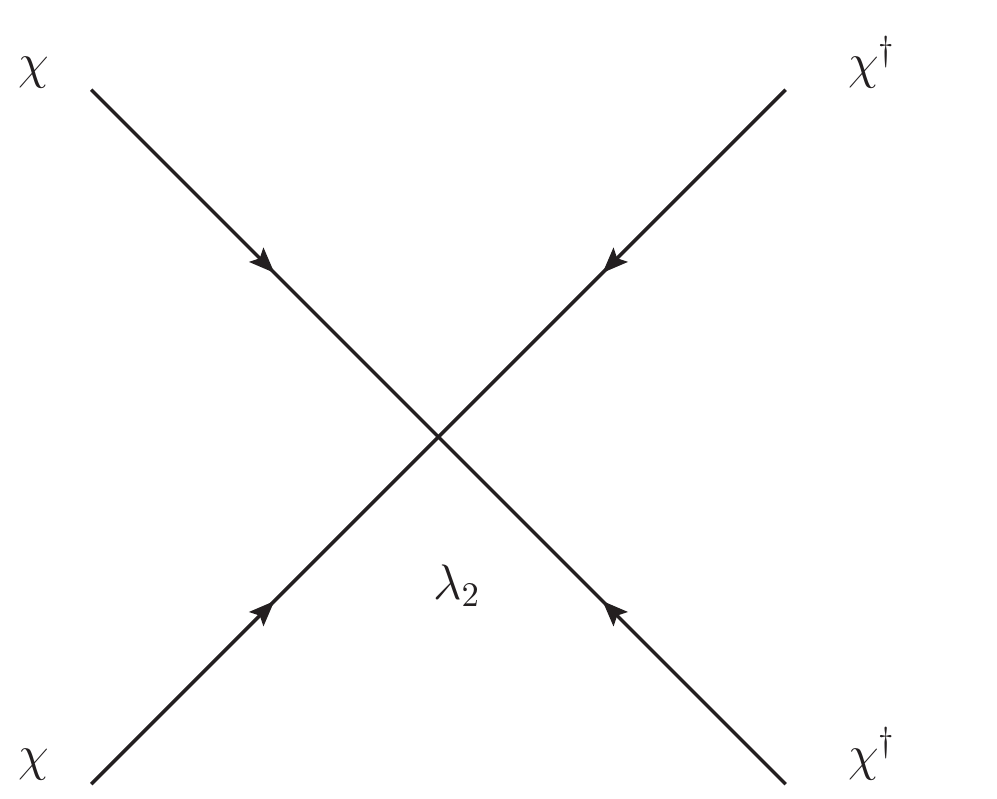}
\includegraphics[scale=0.44]{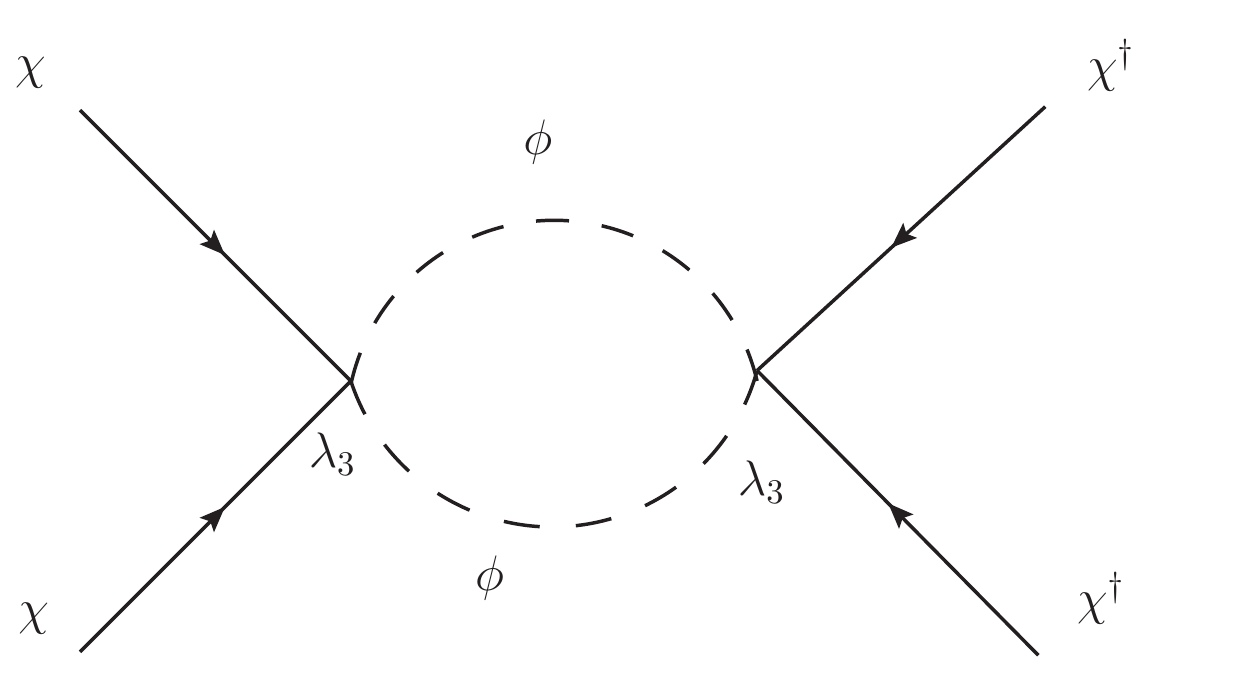}
\includegraphics[scale=0.44]{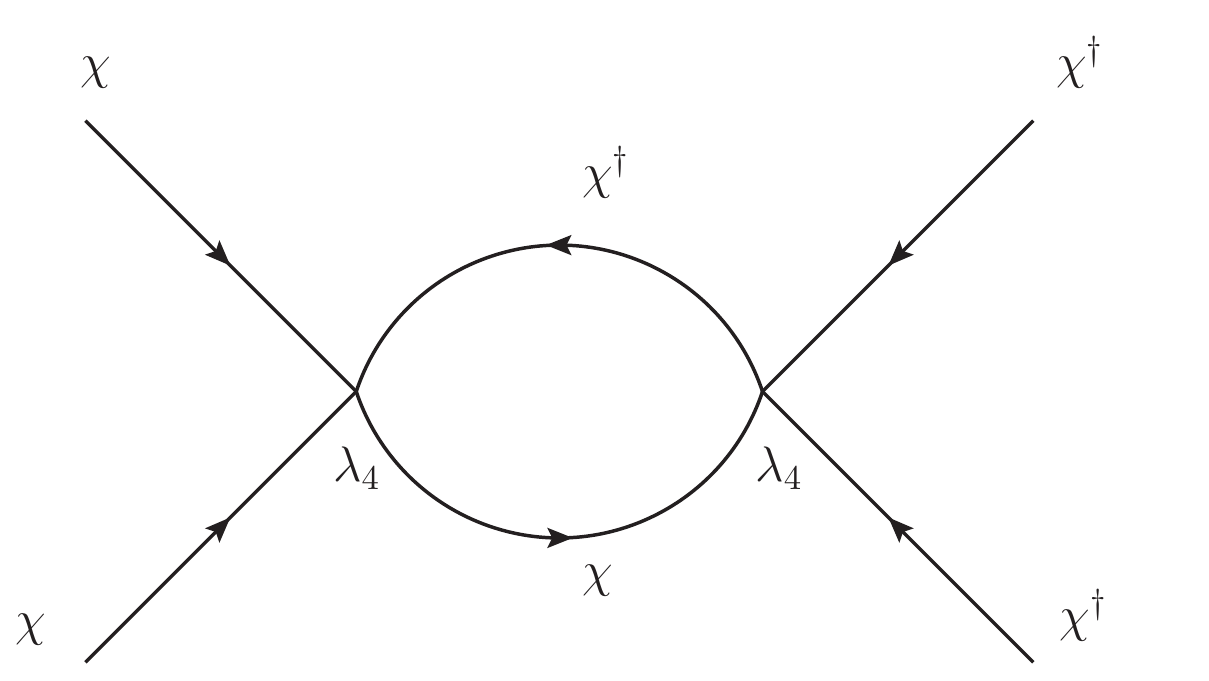}
\caption{\small{\em Relevant tree and one-loop level Feynman diagrams for the CP-violating self-scattering process $\chi+\chi \rightarrow \chi^{\dagger} + \chi^{\dagger}$.}}
\label{fig:diag1}
\end{figure}

The relevant tree and one-loop level Feynman diagrams for the self-scattering process $\chi+\chi \rightarrow \chi^{\dagger} + \chi^{\dagger}$ are shown in Fig.~\ref{fig:diag1}. 
The CP-violation in this process arises from the interference of the tree level and one-loop graphs shown in Fig.~\ref{fig:diag1}. 
There are additional diagrams for this process at the one-loop level at the same order in the couplings. However, those diagrams do not give rise to any CP-violating imaginary parts. Therefore, we have restricted ourselves to the diagrams shown in Fig.~\ref{fig:diag1}. The resulting difference in the matrix elements squared of the two CP-conjugate processes is given by
\begin{align}
|M|^2_{\chi\chi\rightarrow \chi^{\dagger}\chi^{\dagger}}-|M|^2_{\chi^{\dagger}\chi^{\dagger}\rightarrow \chi\chi}=4\,\left[\rm Im(\lambda^*_2\lambda^2_3)\,Im\, I_{\phi\phi}+\,\rm Im(\lambda^*_2\lambda^2_4)\,Im\, I_{\chi^{\dagger}\chi}\right].
\label{eq:ME1}
\end{align}
The imaginary part of the loop amplitudes are non-zero when internal lines in the loop go onshell, and using the Cutkosky rules~\cite{Peskin:1995ev} we obtain them to be
\begin{align}
\rm Im\, I_{\phi\phi}=\dfrac{\beta_\phi}{32 \pi},\hspace{0.5cm} Im\, I_{\chi^{\dagger}\chi}=\dfrac{\beta_\chi}{16 \pi}.
\label{eq:imaginary}
\end{align}
Here, the factor of $\beta_f = (1-4m^2_f/s)^{1/2}$ comes from the two-body phase-space integral, with the usual Mandelstam variable $s=(p_1+p_2)^2$, where $p_1$ and $p_2$ are the four-momenta of the two initial state particles. Since we take $m_\phi << m_\chi$, we can approximate $\beta_\phi \sim 1$. 

With the amplitudes in hand, we can now compute the thermally averaged symmetric rate of this self-scattering process as follows:
\begin{align}
\expval{\sigma v}_1 = \dfrac{|\lambda_2|^2}{16\pi m^2_\chi}\dfrac{1}{\sqrt{\pi x}},
\end{align}
where we have used the partial wave expansion of the annihilation rate, keeping only the leading term in the non-relativistic limit.
As mentioned in the previous sections, and as we can see from this expression, the leading term in the self-scattering rate is proportional to the DM velocity in the thermal bath. Therefore, in the non-relativistic regime, this rate may be suppressed. It is sufficient to consider only the tree-level amplitude for computing the leading contribution to the symmetric annihilation rate. In obtaining the thermal average, we have included the final state symmetry factors here, while the initial state symmetry factors have already been taken into account in the Boltzmann equations.

We can, similarly, obtain the thermal average of the asymmetric reaction rate, which stems from the interference of the tree level and one-loop level amplitudes. The resulting expression is given as
\begin{align}
\expval{\epsilon \sigma v}_1 &= \dfrac{1}{(16\pi m_\chi)^2}\left[\dfrac{1}{\sqrt{\pi x}} \text{Im}(\lambda^*_2\lambda^2_3)+\dfrac{3}{2 x} \text{Im}(\lambda^*_2\lambda^2_4)\right].
\label{eq:eps_sigmav_1}
\end{align}
Here again, the leading term is proportional to the DM velocity, whereas the sub-leading term is proportional to the square of the DM velocity. With this, the effective CP-violation parameter can be obtained as follows: 
\begin{equation}
\epsilon^{1}_{eff} = \expval{\epsilon \sigma v}_1/\expval{\sigma v}_1.
\end{equation}
For $\mathcal{O}(1)$ values of the absolute values of $\lambda_i$, and the phase of the complex coupling combination $\lambda_2^* \lambda_3^2$ (denoted by $\theta_{23}$) set to be $\pi/2$, we obtain for $x>>1$:
\begin{align}
\epsilon^{1}_{eff} \simeq  \dfrac{1}{16\pi} , \,\,\,\, \text{for}\,\, x >> 1.
\label{epsestimate}
\end{align}
Thus for $\mathcal{O}(1)$ couplings, $\epsilon^{1}_{eff} $ is expected to be small, around $0.02$, but can of course be made larger with larger couplings.

\section*{Process 2: Self-scattering $\chi+\chi \rightarrow \chi^{\dagger} + \chi$}

\begin{figure}[htb!]
\centering
\includegraphics[scale=0.39]{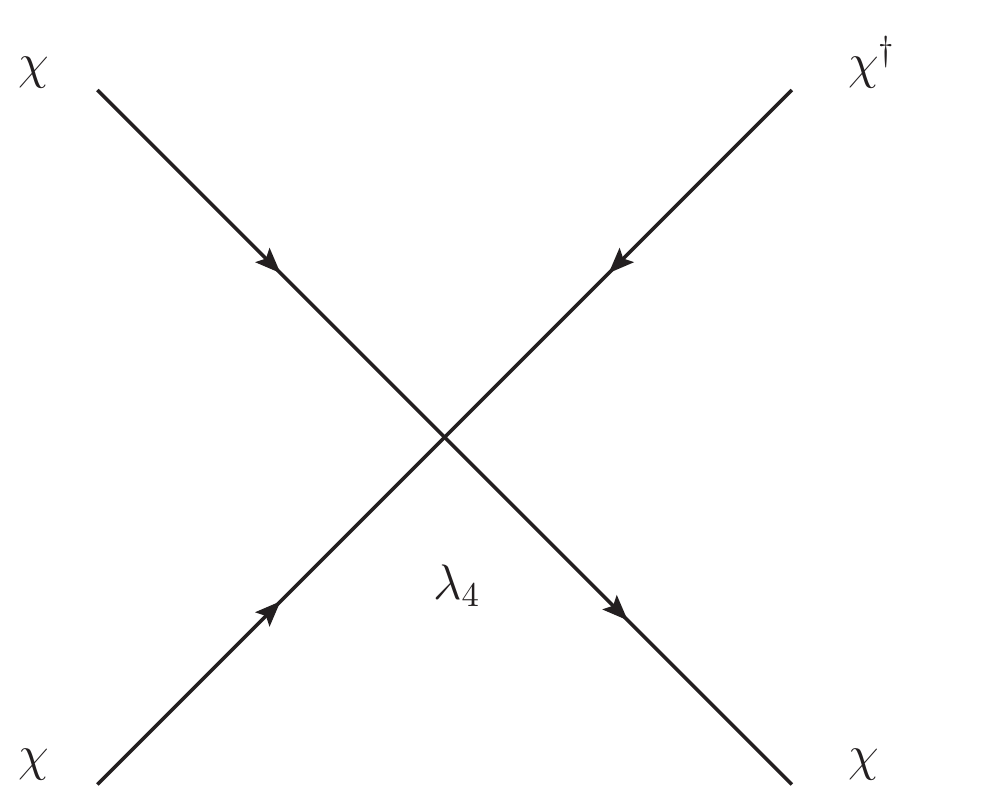}
\includegraphics[scale=0.44]{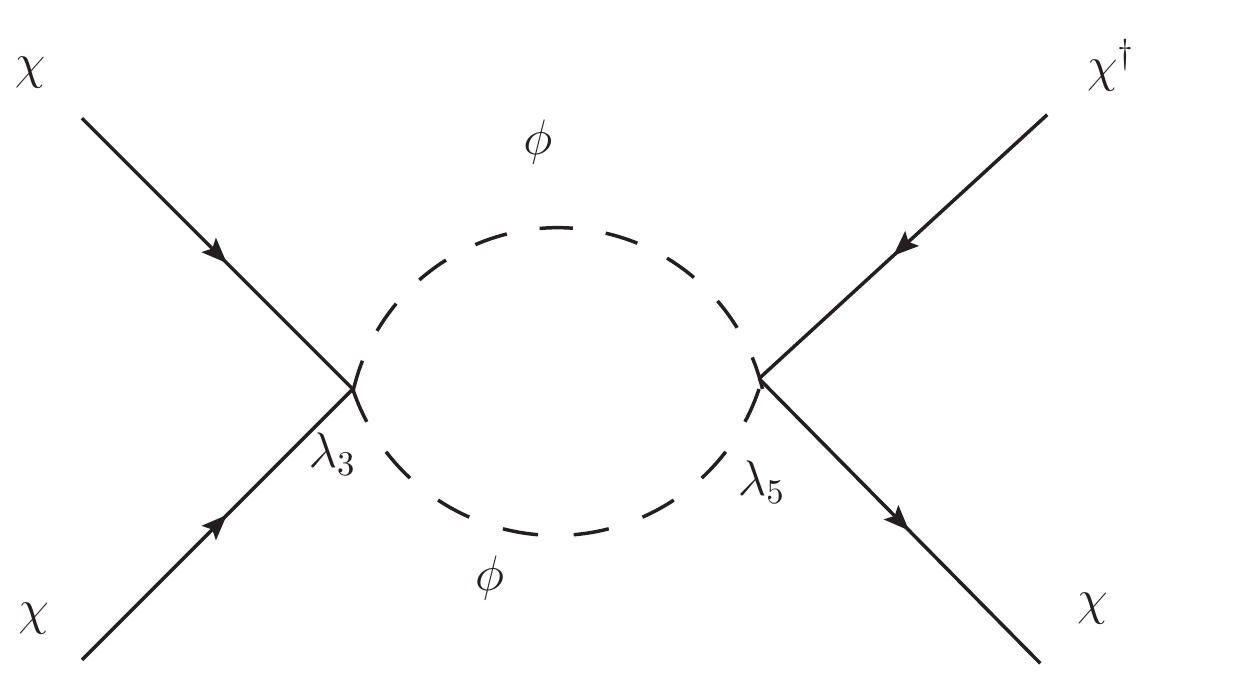}
\includegraphics[scale=0.44]{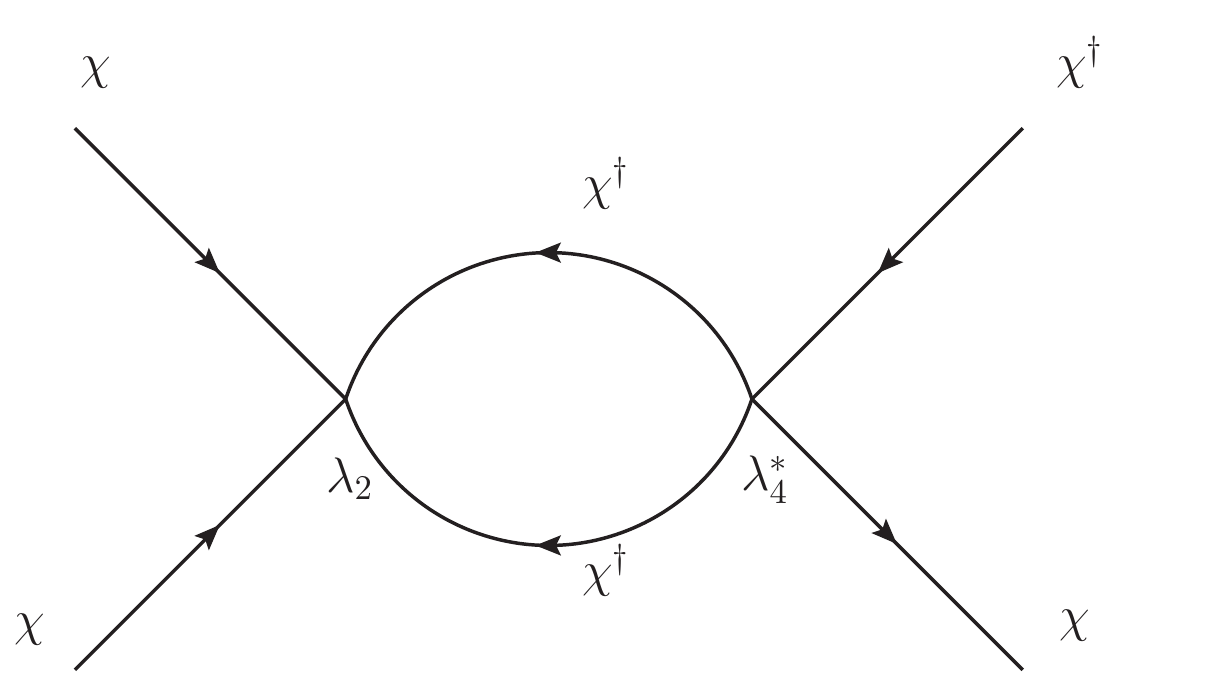}
\caption{\small{\em Relevant tree and one-loop level Feynman diagrams for the CP-violating self-scattering process $\chi+\chi \rightarrow \chi^{\dagger} + \chi$.}}
\label{fig:diag2}
\end{figure}

The second CP-violating self-scattering process, $\chi+\chi \rightarrow \chi^{\dagger} + \chi$ shows very similar behaviour as far as the reaction rates are concerned, except the fact that different coupling combinations appear. In particular, the coupling $\lambda_5$, which controls the rate of CP-conserving $\chi+\chi^{\dagger} \rightarrow \phi + \phi$ annihilation, now appears in the dominant loop contribution to the $\chi+\chi \rightarrow \chi^{\dagger} + \chi$ process, as can be seen from the Feynman diagrams in Fig.~\ref{fig:diag2}. This makes the rate of CP-violation in this process and the rate of CP-conserving annihilations related. 
We can now similarly compute the relevant difference between the squared matrix elements from the interference of the graphs in Fig.~\ref{fig:diag2}
\begin{align}
|M|^2_{\chi\chi\rightarrow \chi^{\dagger}\chi}-|M|^2_{\chi^{\dagger}\chi^{\dagger}\rightarrow\chi^{\dagger}\chi}=4\,\left[\rm Im(\lambda_2\lambda^{*2}_4)\,Im\, I_{\chi^{\dagger}\chi^{\dagger}}+\,\rm Im(\lambda^*_4\lambda_3)\lambda_5\,Im\, I_{\phi\phi}\right],
\label{eq:ME2}
\end{align}
with the following imaginary part of the new loop amplitude, in addition to the ones shown in Eq.~\ref{eq:imaginary}:
\begin{align}
\rm Im\, I_{\chi^{\dagger}\chi^{\dagger}}=\dfrac{\beta_\chi}{32 \pi}.
\end{align}
The thermal averaged symmetric annihilation rate shows the same velocity dependence as before, and is given by
\begin{align}
\expval{\sigma v}_2 = \dfrac{|\lambda_4|^2}{8\pi m^2_\chi}\dfrac{1}{\sqrt{\pi x}}\hspace{1cm}.
\end{align}
Similarly the $x-$dependence of the asymmetric annihilation rate is also similar, but now with the explicit appearance of the $\lambda_5$ coupling
\begin{align}
\expval{\epsilon \sigma v}_2 &= \dfrac{1}{(16\pi m_\chi)^2}\left[\dfrac{3}{2 x} \text{Im}(\lambda_2\lambda^{*2}_4)+\dfrac{2}{\sqrt{\pi x}} \text{Im}(\lambda^*_4\lambda_3)\lambda_5\right].
\label{eq:eps_sigmav_2}
\end{align}
Thus, the CP-violation will proportionally increase if we increase the pair-annihilation rate $\expval{\sigma v}_A$. We can also write the effective CP-violation parameter as 
\begin{align}
\epsilon^{2}_{eff} \simeq \dfrac{\lambda_5}{16\pi} , \,\,\,\, \text{for}\,\, x >> 1.
\end{align}
Here, as before, we have set $|\lambda_i|=\mathcal{O}(1)$, except for $\lambda_5$ which is kept variable, and all the effective phases to be $\pi/2$. We shall see in the following that to obtain a large asymmetry in the DM sector, one generally requires a large rate of $\expval{\sigma v}_A$, and hence a large $\lambda_5$. In such cases, we may have $\epsilon^{2}_{eff} \simeq 0.25$, for $\lambda_5 \sim \mathcal{O}{(4\pi)}$.

\section*{Process 3: CP-violating Annihilation $\chi+\chi \rightarrow \phi + \phi$}
\begin{figure}[htb!]
\centering
\includegraphics[scale=0.39]{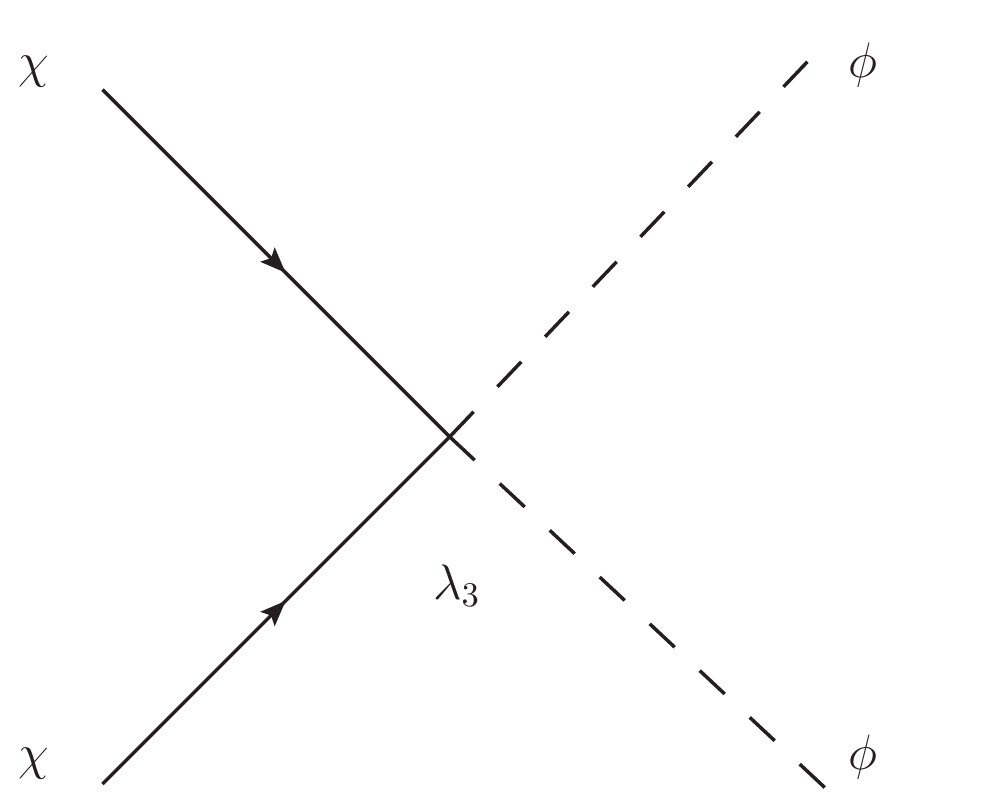}
\includegraphics[scale=0.43]{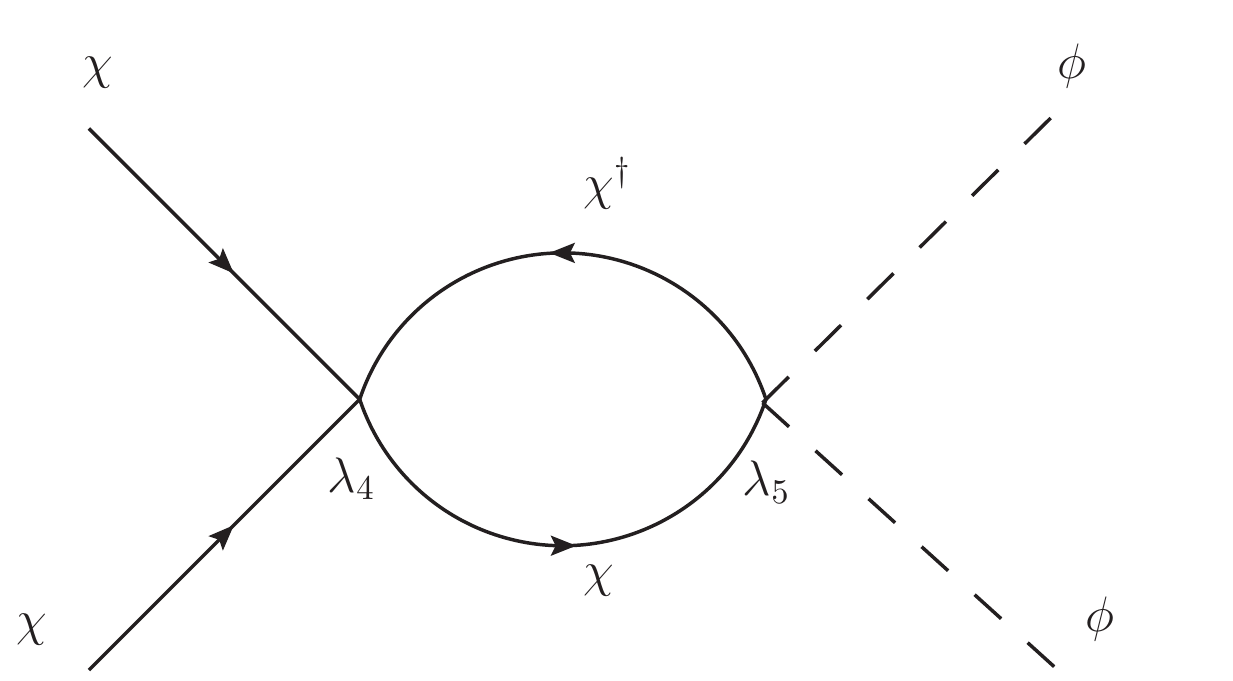}
\includegraphics[scale=0.43]{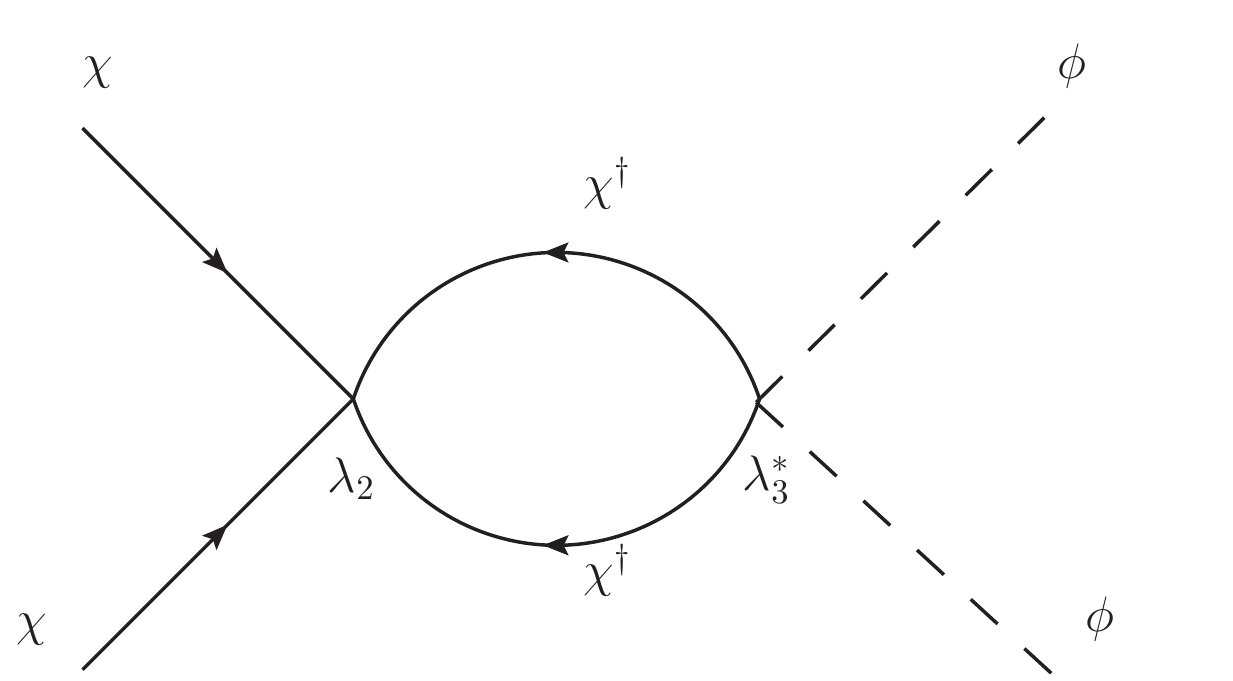}
\caption{\small{\em Relevant tree and one-loop level Feynman diagrams for the CP-violating annihilation process $\chi+\chi \rightarrow \phi + \phi$.}}
\label{fig:diag3}
\end{figure}
Finally, we come to the third CP-violating process, which is an annihilation that reduces the total DM and anti-DM number as well, unlike the previous two self-scattering processes. The relevant Feynman diagrams are shown in Fig.~\ref{fig:diag3}. One of the loop diagrams here is again proportional to the CP-conserving $\lambda_5$ coupling. The difference between the matrix elements squared of the CP-conjugate processes is
\begin{align}
|M|^2_{\chi\chi\rightarrow \phi\phi}-|M|^2_{\chi^{\dagger}\chi^{\dagger}\rightarrow\phi\phi}=4\,\left[\rm Im(\lambda^*_3\lambda_4)\lambda_5\,Im\, I_{\chi^{\dagger}\chi}+\,\rm Im(\lambda_2\lambda^{*2}_3)\,Im\, I_{\chi^{\dagger}\chi^{\dagger}}\right]
\label{eq:ME3}
\end{align}
Unlike in the self-scatterings, the thermally averaged symmetric scattering rate is no longer velocity suppressed in the leading term, and is given as
\begin{align}
\expval{\sigma v}_3 = \dfrac{|\lambda_3|^2}{32\pi m^2_\chi}\left(1+\dfrac{3}{4x}\right).
\end{align}
The asymmetric scattering rate also shows a somewhat different behaviour, with both terms in the thermal average now suppressed by a single power of the DM velocity:
\begin{align}
\expval{\epsilon \sigma v}_3 &= \dfrac{1}{(16\pi m_\chi)^2}\dfrac{1}{\sqrt{\pi x}}\bigg[\text{Im}(\lambda_2\lambda^{*2}_3)+ 2\, \text{Im}(\lambda^*_3\lambda_4)\lambda_5 \bigg].
\label{eq:eps_sigmav_3}
\end{align}
Thus, the effective CP-violation parameter, for $|\lambda_i| \sim \mathcal{O}(1)$ and the phases in the relevant coupling combinations set to $\pi/2$, now carries an explicit $x-$dependence even in the leading term:
\begin{align}
 \epsilon^{3}_{eff} \simeq -\dfrac{1}{8\pi}\dfrac{1+2\lambda_5}{\sqrt{\pi x}} , \,\,\,\, \text{for}\,\, x >> 1.
\end{align}

With all our computations in place, we can check the unitarity sum rules explicitly upto the next-to-leading order in perturbation theory. First of all, using Eqs.~\ref{eq:ME1},~\ref{eq:ME2} and~\ref{eq:ME3}, we see that the sum rule in Eq.~\ref{eq:eps_unitarity} holds, namely
\begin{equation}
\int dPS_2 \,\epsilon_1 \,|M_0|^2_1 + \int dPS_2 \,\epsilon_2 \,|M_0|^2_2 + \int dPS_2 \,\epsilon_3 \,|M_0|^2_3 = 0.
\end{equation}
The same unitarity sum rule can also be written in terms of the thermally averaged asymmetric reaction rates:
\begin{align}
\expval{\epsilon \sigma v}_1 + \expval{\epsilon \sigma v}_2 + \expval{\epsilon \sigma v}_3 =0,
\label{eq:unitarity_sum_epsilon}
\end{align}
which we can again easily check at this order in perturbation theory using the expressions in Eqs.~\ref{eq:eps_sigmav_1}, \ref{eq:eps_sigmav_2} and \ref{eq:eps_sigmav_3}.

We now point out a key observation of our study. If the CP-violating annihilation process is absent, i.e., if we set $\lambda_3=0$, then we can see from Eq.~\ref{eq:unitarity_sum_epsilon} that $\expval{\epsilon \sigma v}_1 + \expval{\epsilon \sigma v}_2=0$. Equivalently, this can be verified also by setting $\lambda_3=0$ in Eqs.~\ref{eq:eps_sigmav_1} and \ref{eq:eps_sigmav_2}. And therefore, by Eq.~\ref{eq:asym}, the asymmetric yield will not be generated, starting from a symmetric initial condition. This is because $\expval{\epsilon \sigma v}_S(=\expval{\epsilon \sigma v}_1 + \expval{\epsilon \sigma v}_2)$ acts as the source term for the asymmetry in our parametrization of the Boltzmann equations, as seen from both Eq.~\ref{eq:asym}, and the approximate analytic solution in Eq.~\ref{eq:analytic2}. Thus we see that the presence of only the CP-violating self-scatterings cannot generate the asymmetry. Similarly, only the presence of the CP-violating annihilation cannot generate an asymmetry either, by the unitarity sum rule. It is the simultaneous presence of both the self-scatterings and the annihilation that can lead to a generation of particle-antiparticle asymmetries.

A remark about the number of independent phases in the coupling combinations that appear in the $\expval{\epsilon \sigma v}_i$ is in order. Defining each coupling as $\lambda_i =|\lambda_i| e^{i\theta_i}$, we can write the three rates in terms of two relative phase angles, namely $\theta_{23} = 2\theta_3-\theta_2$ and $\theta_{24} = 2\theta_4-\theta_2$. The third relevant angle is $\theta_4-\theta_3=(\theta_{24}-\theta_{23})/2$, and hence is not independent. In what follows, we shall set $\theta_{23}= - {\theta}_{24}=\pi/2$, to maximize the CP-violating effects.

\section*{Process 4: CP-conserving Annihilation $\chi+\chi^\dagger \rightarrow \phi + \phi$}
We have already mentioned the role of the CP-conserving annihilation in indirectly controlling the particle-antiparticle asymmetry in Sec.~\ref{sec:sec3}. In addition to the asymmetry, such a pair annihilation process, if present, also strongly affects the final relic abundance of the DM and anti-DM particles. The thermally averaged annihilation rate for this process in the complex singlet scenario, keeping the first two terms in the partial-wave expansion, is given as follows:
\begin{align}
 \expval{\sigma v}_A = \dfrac{|\lambda_5|^2}{64\pi m^2_\chi}\left(1+\dfrac{3}{4x}\right).
\end{align}

\section*{The $\chi+\phi \rightarrow \chi^{\dagger}+\phi$ process}
As mentioned in Sec.~\ref{sec:sec2}, we can also have the CP-violating DM anti-DM conversion process $\chi+\phi \rightarrow \chi^{\dagger}+\phi$. As shown by the explicit computation described in Appendix~\ref{App.A}, the net contribution to CP-violation from all the Feynman diagrams contributing to this process is 
\begin{align}
\int dPS_2\left(|M|^2_{\chi\phi\rightarrow \chi^\dagger\phi}-|M|^2_{\chi^{\dagger}\phi \rightarrow \chi\phi}\right) \propto{\rm Im(\lambda^*_3\mu_1)} \times {\rm ~Terms ~of}~\mathcal{O}({\hat\mu^3/m_\chi^4}),
\end{align}
where, $\hat{\mu}$ is either $\mu$, $\mu_1$ or $\mu_\phi$. This is in contrast to the CP-violation stemming from the processes discussed in Sec.~\ref{sec:sec2}, which is independent of the dimensionful couplings. Since we have assumed $\mu_1 \simeq 0$ in order to avoid the stringent constraints from spin-independent DM direct detection probes, clearly, the contribution to CP-violation from the $\chi\phi\rightarrow \chi^\dagger\phi$ process is much smaller compared to the ones we have included in the Boltzmann equations for the (anti-)DM number densities.

\subsection{Dark matter density and asymmetry: the interplay of different processes}
\begin{figure}[htb!]
\centering
\includegraphics[scale=0.55]{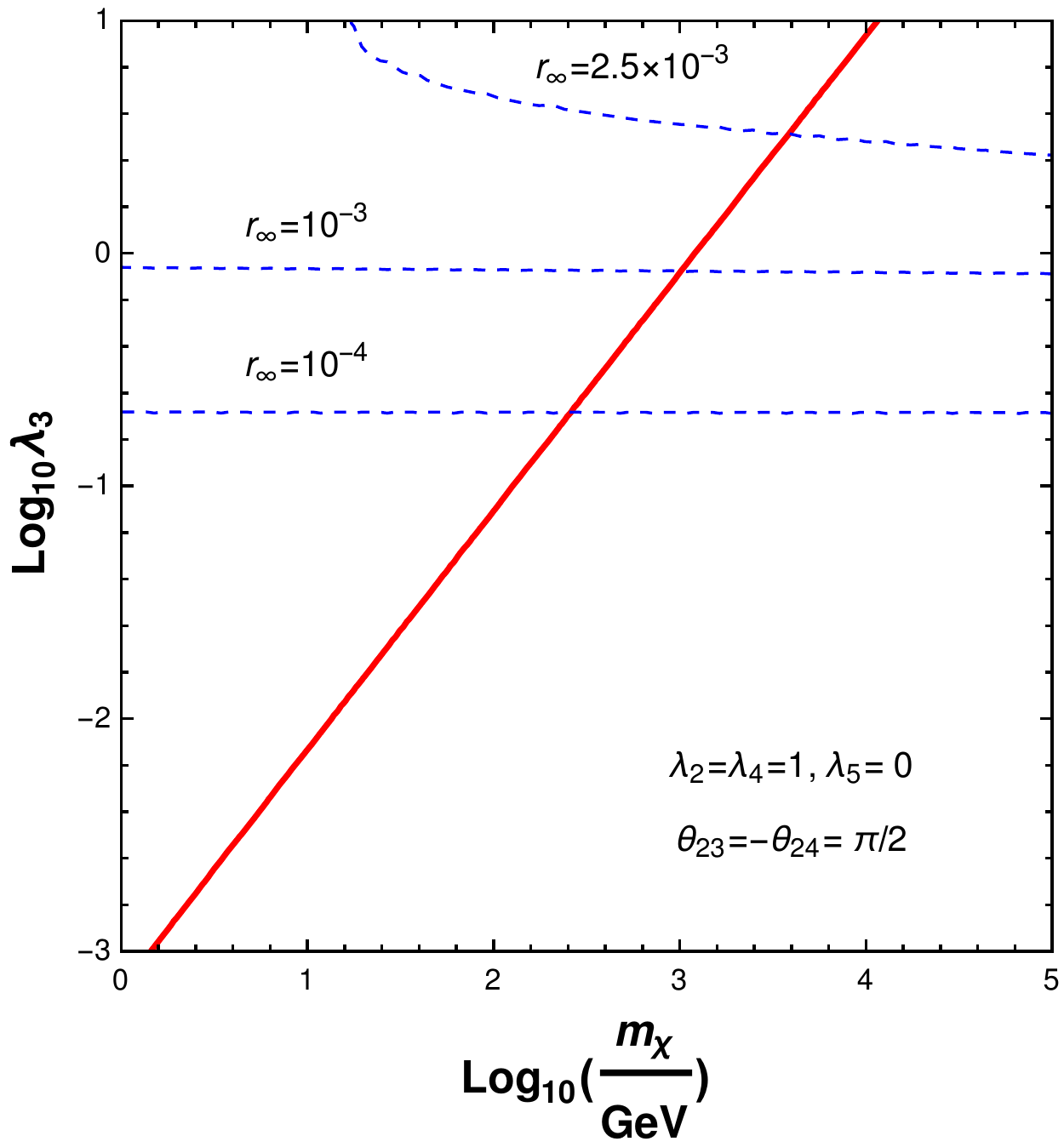}
\caption{\small{\em Contour in the $\lambda_3 - m_\chi$ plane for which the DM and anti-DM densities saturate the DM density of the Universe, with $\Omega h^2 \simeq 0.12$ (red solid line). Also shown are the contours of the constant final asymmetry parameter $r_\infty=\lvert Y_{\Delta\chi} \rvert / Y_S$ (blue dashed lines). Here, the coupling $\lambda_3$ determines the symmetric annihilation rate for the $\chi + \chi \rightarrow \phi + \phi$ process, and also the CP-violation through loop effects. For this figure, we have set the CP-conserving pair-annihilation coupling $\lambda_5=0$.}}
\label{fig:mass_lambda3}
\end{figure}  

Having determined the CP-violation and the symmetric and asymmetric annihilation rates in the complex scalar singlet model, we can now compute the relic abundance of the DM and anti-DM system. Due to the possible CP-violation, there can also be a resulting particle-antiparticle asymmetry in the DM sector. We have defined the measure of the final particle-antiparticle asymmetry parameter as $r_\infty=\lvert Y_{\Delta\chi} \rvert / Y_S$, where the asymptotic values of the yields are considered. Clearly, we have $0 \leq r_\infty \leq 1$. Here, $r_\infty=0$ corresponds to the completely symmetric limit, in which the asymptotic yields of the DM and anti-DM are the same. On the other hand, $r_\infty=1$ corresponds to the completely asymmetric limit, in which only either the DM or the anti-DM species survives. As we shall see in the following, depending upon the dominance of either the CP-conserving or the CP-violating pair-annihilation couplings, we obtain two different regimes for the DM $-$ a nearly symmetric regime, and an asymmetric regime.

\begin{figure}[htb!]
\centering
\includegraphics[scale=0.56]{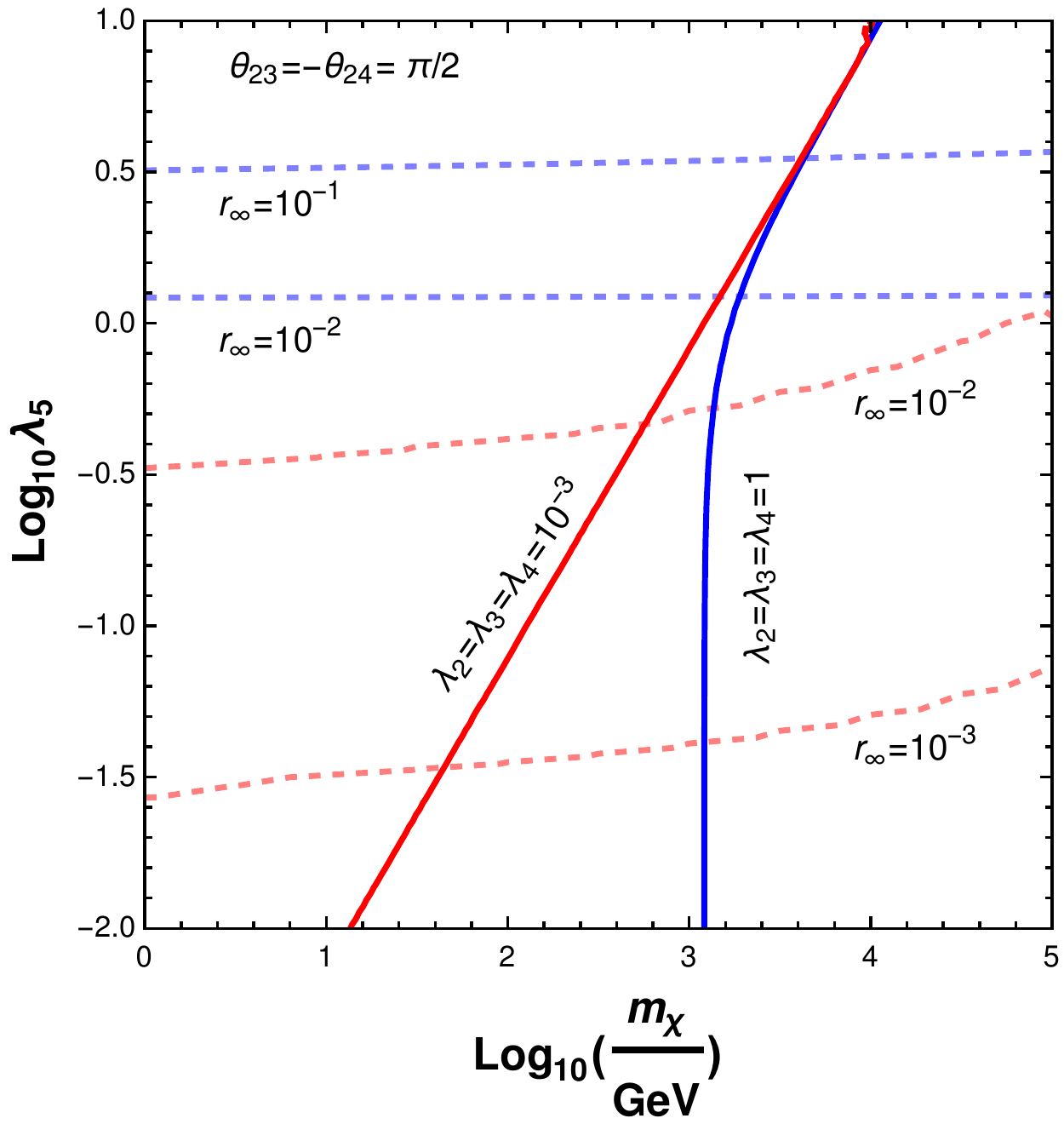}
\caption{\small{\em Contours in the $\lambda_5 - m_\chi$ plane for which the DM and anti-DM densities saturate the DM density of the Universe, with $\Omega h^2 \simeq 0.12$ (blue and red solid lines), for two different choices of $\lambda_2,\lambda_3$ and $\lambda_4$, as indicated against the lines. Also shown are the colour-coded contours of the constant final asymmetry parameter $r_\infty=\lvert Y_{\Delta\chi} \rvert / Y_S$ (with the blue and red  dashed lines corresponding to the scenarios with $\lambda_2=\lambda_3=\lambda_4=1$ and $\lambda_2=\lambda_3=\lambda_4=10^{-3}$, respectively). Here, the coupling $\lambda_5$ determines the CP-conserving pair annihilation rate for the $\chi + \chi^\dagger \rightarrow \phi + \phi$ process, and also the CP-violation through loop effects.}}
\label{fig:mass_lambda_5}
\end{figure}

To begin with, let us consider the range of DM mass that is allowed in this scenario. We find that the DM relic abundance required for saturating the observed DM density of $\Omega h^2 \simeq 0.12$~\cite{Aghanim:2018eyx} can be obtained for $m_\chi$ in the range from $\mathcal{O}({\rm GeV})$ to $\mathcal{O}(10 {~\rm  TeV})$, with the relevant couplings kept within their perturbative limits. Since both $\lambda_3$ and $\lambda_5$ can play a dominant role in determining the DM relic density, as seen in Sec.~\ref{sec:sec3} and in particular in Eqs.~\ref{eq:YS_approx_1} and \ref{eq:YS_approx_2}, we show the allowed DM mass range as a function of these two couplings separately in Figs.~\ref{fig:mass_lambda3} and ~\ref{fig:mass_lambda_5}. 

\begin{figure}[htb!]
\centering
\includegraphics[scale=0.6]{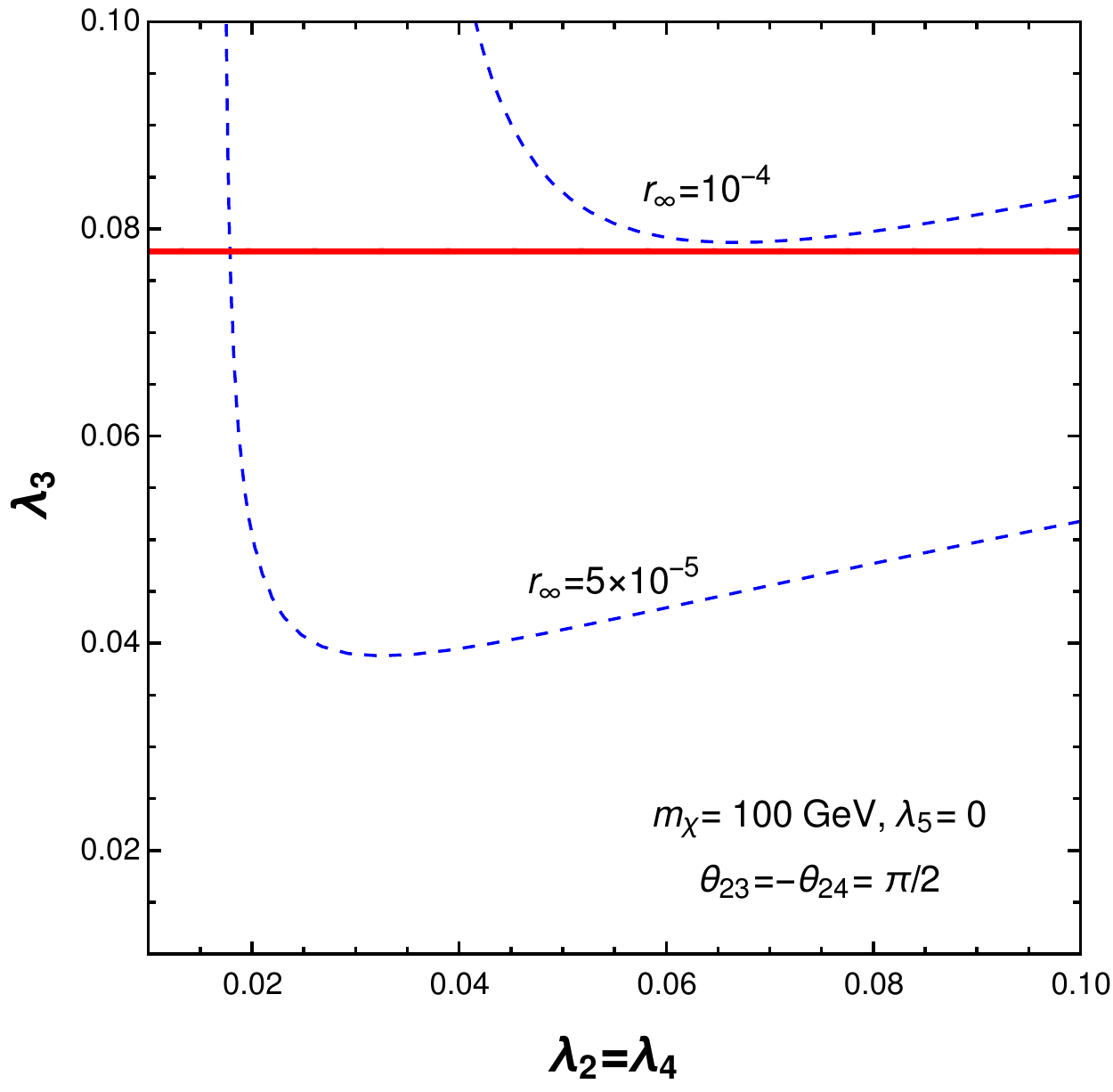}
\caption{\small{\em Contour in the $\lambda_3 - \lambda_{2,4}$ plane  for which the DM and anti-DM densities saturate the DM density of the Universe, with $\Omega h^2 \simeq 0.12$ (red solid line). Also shown are the contours of the constant final asymmetry parameter $r_\infty=\lvert Y_{\Delta\chi} \rvert / Y_S$ (blue dashed lines). We have set the two coupling parameters determining the DM self-scattering rates to be equal ($\lambda_2 = \lambda_4$) for simplicity. The CP-conserving coupling has been fixed at $\lambda_5=0$, see text for details.}}
\label{fig:results_wthout_ann}
\end{figure}

In Fig.~\ref{fig:mass_lambda3}, we show the contour in the $\lambda_3 - m_\chi$ plane for which the DM and anti-DM densities furnish $\Omega h^2 \simeq 0.12$ (red solid line). We also show the contours of the constant final asymmetry parameter $r_\infty$ defined above (blue dashed lines). For this figure, we have fixed the DM self-couplings as $\lambda_2=\lambda_4=1$, and have
set the CP-conserving pair-annihilation coupling $\lambda_5=0$. The reason for setting $\lambda_5=0$ is to demonstrate the role of $\lambda_3$ in determining the density. The effective CP-violating phases are set to $\pi/2$. As we can see from this figure, increasing $\lambda_3$ implies a corresponding increase in the mass $m_\chi$ that saturates the DM abundance. The final asymmetry parameter $r_\infty$ is found to be rather small, of the order of $10^{-4}$ to $10^{-3}$, indicating a nearly symmetric DM. This is primarily because with these choices of parameters the CP-violation is not large, and with the CP-conserving pair-annihilation switched off, the symmetric part is not subsequently removed as well. In addition to that, in this scenario, the asymmetric reaction rate for the CP-violating processes are also reduced, as seen in, for example, Eq.~\ref{eq:eps_sigmav_2}.

How do the DM density and the asymmetry parameter vary as a function of the pair-annihilation ($\lambda_3$) and self-scattering couplings ($\lambda_2, \lambda_4$)? We show this correlation in Fig.~\ref{fig:results_wthout_ann}. For this figure, we have kept the DM mass fixed at $m_\chi=100$ GeV, and continue to consider $\lambda_5=0$. As we can see, the DM asymmetry continues to be small, and in the parameter range of interest can vary from $10^{-5}$ to $10^{-4}$. Furthermore, with such a small asymmetry, the DM self-scatterings play no significant role in deciding the DM density, which is fixed by $\lambda_3$. However, the DM composition varies with the variation of the self-interaction couplings.

We now move to the scenario in which the CP-conserving pair-annihilation coupling is non-zero, and potentially large. We find that in such a case, a larger asymmetry is obtained. Furthermore, we find a novel and interesting effect that the self-scattering couplings now play an equally dominant role in deciding both the DM  density and asymmetry, as do the annihilation couplings. We first show the contours in the $\lambda_5 - m_\chi$ plane for which the DM and anti-DM densities saturate the DM density of the Universe in Fig.~\ref{fig:mass_lambda_5}. The mass range obtained is again similar to the one obtained in Fig.~\ref{fig:mass_lambda3}. For the blue solid line, we have fixed $\lambda_2=\lambda_3=\lambda_4=1$, while for the red solid line we have set $\lambda_2=\lambda_3=\lambda_4=10^{-3}$, for showing two illustrative cases. We see that when $\lambda_3=1$, there is no significant variation in the DM density for $\lambda_5 < 1$, since $\lambda_3$ also plays a dominant role in determining it. However, for $\lambda_5>\lambda_3$, the density changes rapidly with varying  $\lambda_5$, as seen from the $\lambda_5 > 1$ region along the blue line, and the entire region along the red line. 

\begin{figure}[htb!]
\centering
\includegraphics[scale=0.59]{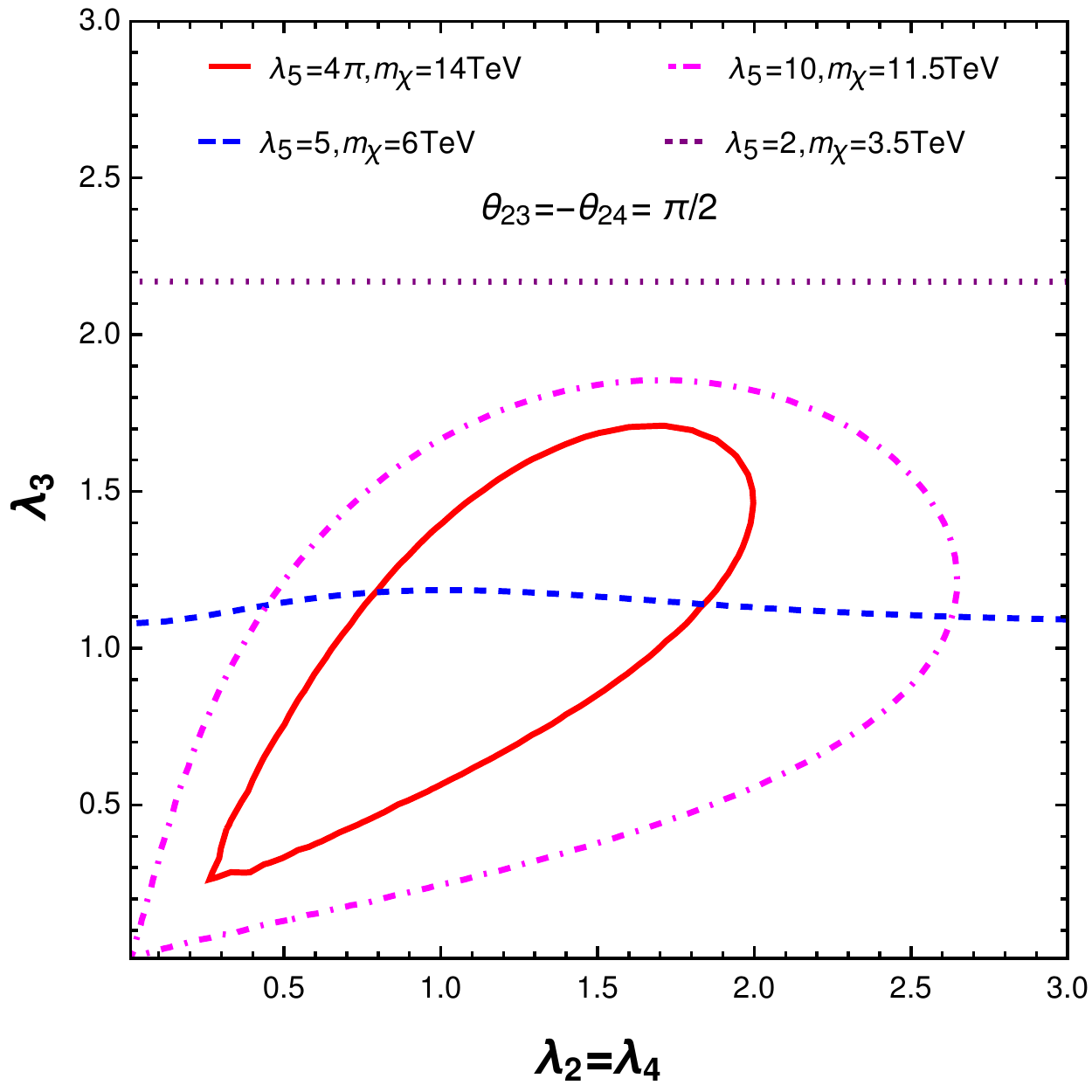}
\includegraphics[scale=0.59]{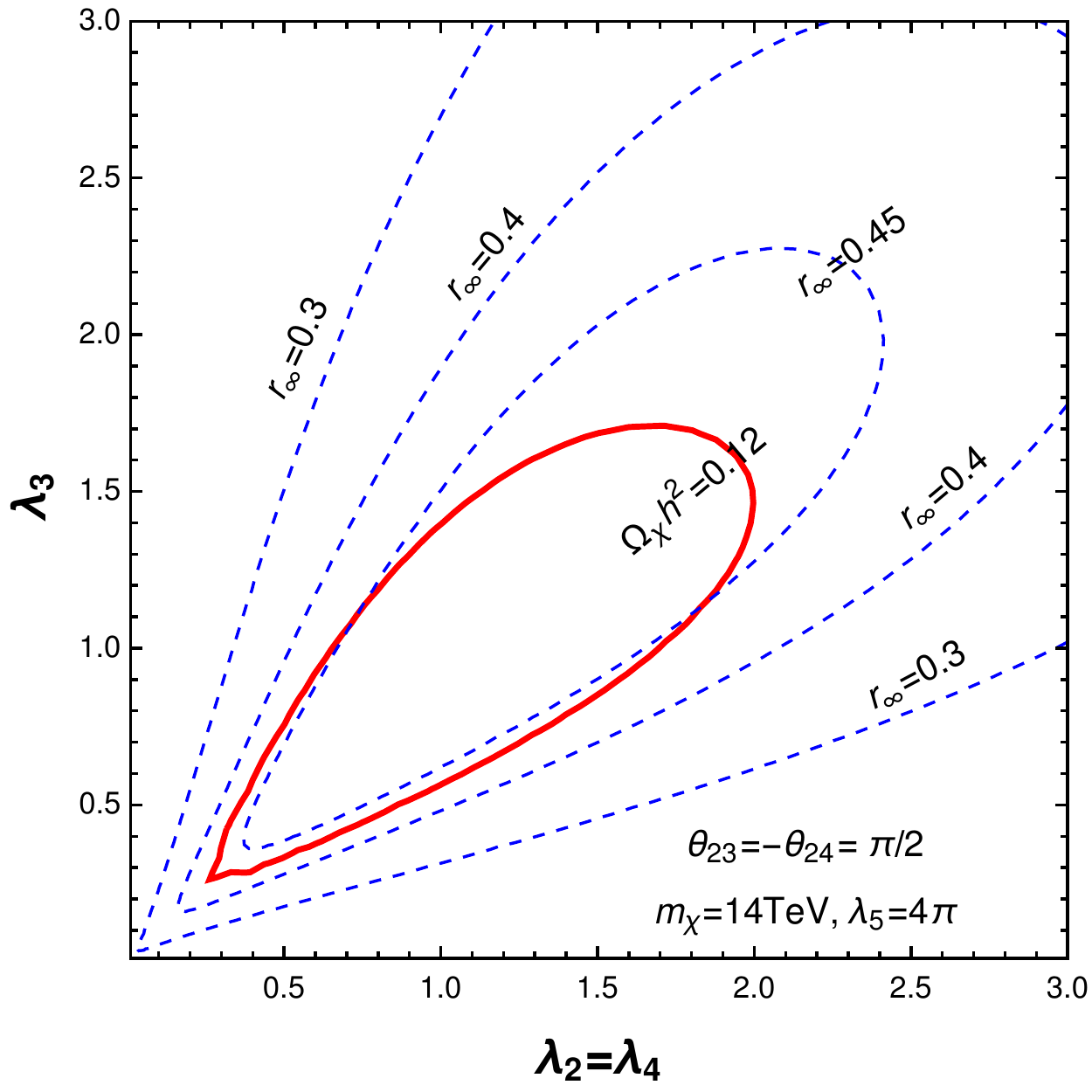}
\caption{\small{\em (Left Panel:) Contours in the $\lambda_3 - \lambda_{2,4}$ plane  for which the DM and anti-DM densities saturate the DM density of the Universe, with $\Omega h^2 \simeq 0.12$, for different values of the CP-even coupling $\lambda_5$ and DM mass $m_\chi$. (Right Panel:) Contours of the constant final asymmetry parameter $r_\infty=\lvert Y_{\Delta\chi} \rvert / Y_S$ (blue dashed lines), with a representative value of $\lambda_5=4 \pi$ and DM mass $m_\chi=14$ TeV. We have set the two coupling parameters determining the DM self-scattering rates to be equal ($\lambda_2 = \lambda_4$) for simplicity. In both the figures, we clearly see the strong interplay of the DM annihilation and self-scattering processes in determining both the DM density and asymmetry.}}
\label{fig:results_with_ann}
\end{figure}

With the introduction of the CP-conserving pair annihilation, the asymmetry parameter $r_\infty$ is seen to take values of the order of $10^{-1}$ (or higher, as we shall see in the following). Thus we can obtain a large asymmetry in the DM sector in this scenario. As discussed earlier, this is due to the fact that after the CP-violating processes decouple, thereby freezing out the difference in the DM and anti-DM yields $Y_{\Delta \chi}$, the CP-conserving pair annihilation processes with larger rates can still remain active. These pair annihilations at later epochs reduce the sum of the DM and anti-DM yields $Y_S$. This leads to the enhancement of the final asymmetry parameter $r_\infty = \lvert Y_{\Delta\chi} \rvert / Y_S$. The $r_\infty$ contours shown in Fig.~\ref{fig:mass_lambda_5} are colour coded, with the blue dashed contours corresponding to the $\lambda_2=\lambda_3=\lambda_4=1$ scenario, while the red dashed ones corresponding to the $\lambda_2=\lambda_3=\lambda_4=10^{-3}$ scenario. 

We now study the variation of the DM density and the asymmetry parameter as a function of the pair-annihilation ($\lambda_3$) and self-scattering couplings ($\lambda_2, \lambda_4$), in the presence of the $\lambda_5$ coupling. As shown in the left panel of Fig.~\ref{fig:results_with_ann}, there are strong correlations between the annihilation and self-scattering in such a scenario. We have shown the contours in this plane where the relic abundance is satisfied, with different sets of values for the CP-conserving coupling $\lambda_5$ and the DM mass $m_\chi$. For each choice of $\lambda_5$, we have chosen $m_\chi$ such that the relic density condition can be satisfied in some region in the coupling plane. The correlations are especially pronounced for larger values of $\lambda_5$, as seen from the shown representative values of $\lambda_5=10$ (pink dot-dashed line) and $\lambda_5=4\pi$ (red solid line), both of which are within the limits of perturbation theory, such that our computations can be reliable. There are small correlations for $\lambda_5=5$ as well (blue dashed line), while no observable correlations are found for $\lambda_5=2$ (violet dotted line). The $\lambda_5=2$ scenario is thus similar to the previously discussed case of no CP-conserving annihilation, as both lead to a rather small final asymmetry. 

In order to understand the physics behind these strong correlations, we pick out one representative value of $\lambda_5=4\pi$, and study  the variation of the asymmetry parameter $r_\infty$ in the right panel of Fig.~\ref{fig:results_with_ann}. We see that $r_\infty$, and hence the asymmetry in the DM sector can be large in this scenario, with the contours upto $r_\infty=0.45$ spanning a large range of the parameter space. This explains the reason behind the strong correlations being observed. For with such a large asymmetry, the total DM and anti-DM yield strongly depends upon the asymmetric yield $Y_{\Delta \chi}$ itself, as can be seen from the analytic approximation in Eq.~\ref{eq:YS_approx_2}, i.e., the asymmetry plays a dominant role in fixing the relic density. This is the reason we observe the strong interplay between the DM self-scatterings and the annihilations in fixing not only the DM asymmetry, but the DM relic density itself. This role of the DM self-scatterings is a novel observation, which should be of considerable interest in the study of  DM cosmology. We would like to emphasize again here that within this scenario of complex singlet scalar DM, we cannot switch off the self-scattering processes, but have only the CP-violating annihilations, or vice versa, and generate an asymmetry in the DM sector. The simultaneous presence of both these types of processes is necessary in order to have a non-zero CP-violation, as dictated by the unitarity sum rules. 

\begin{figure}[htb!]
\centering
\includegraphics[scale=0.52]{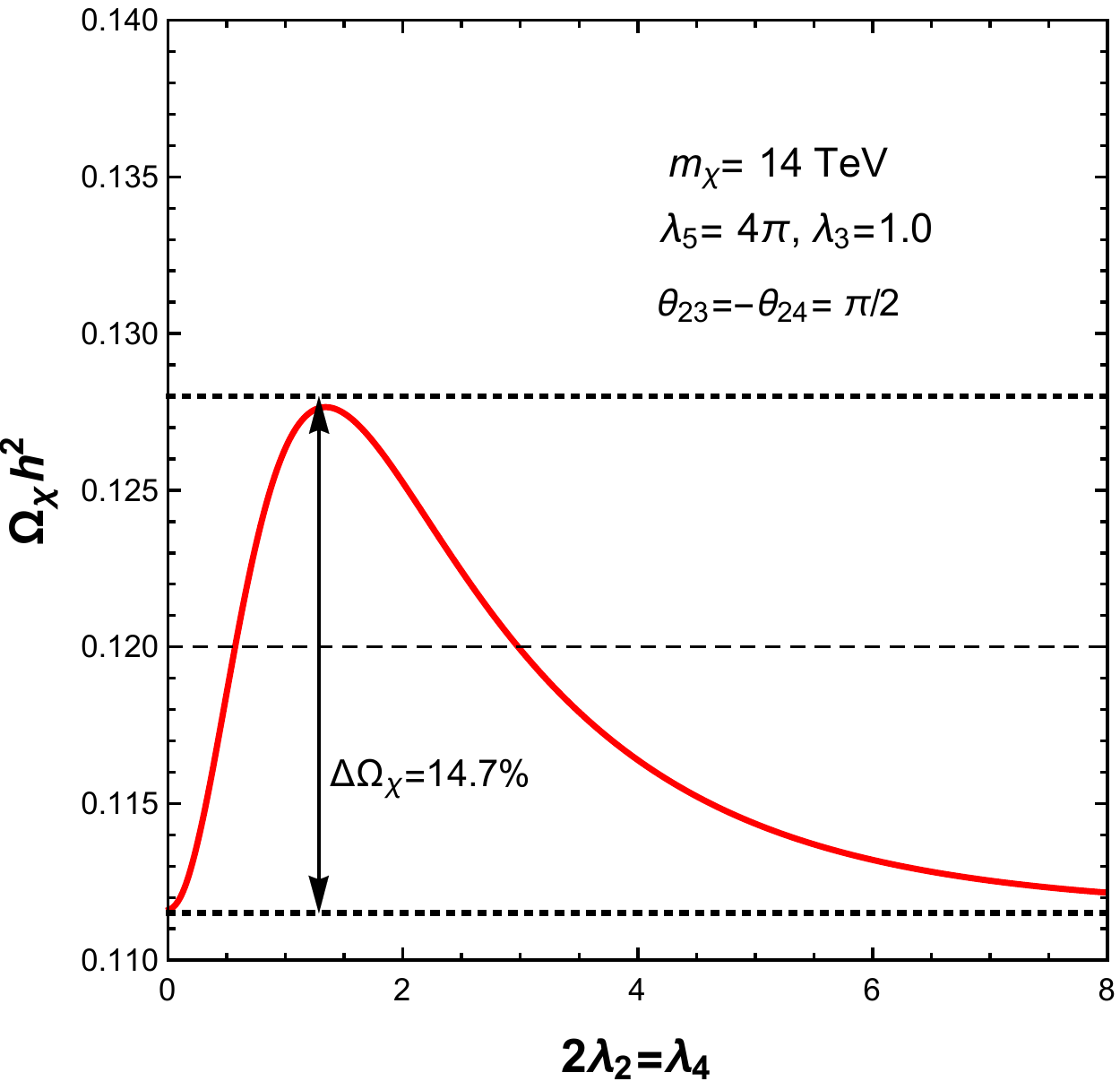}
\includegraphics[scale=0.52]{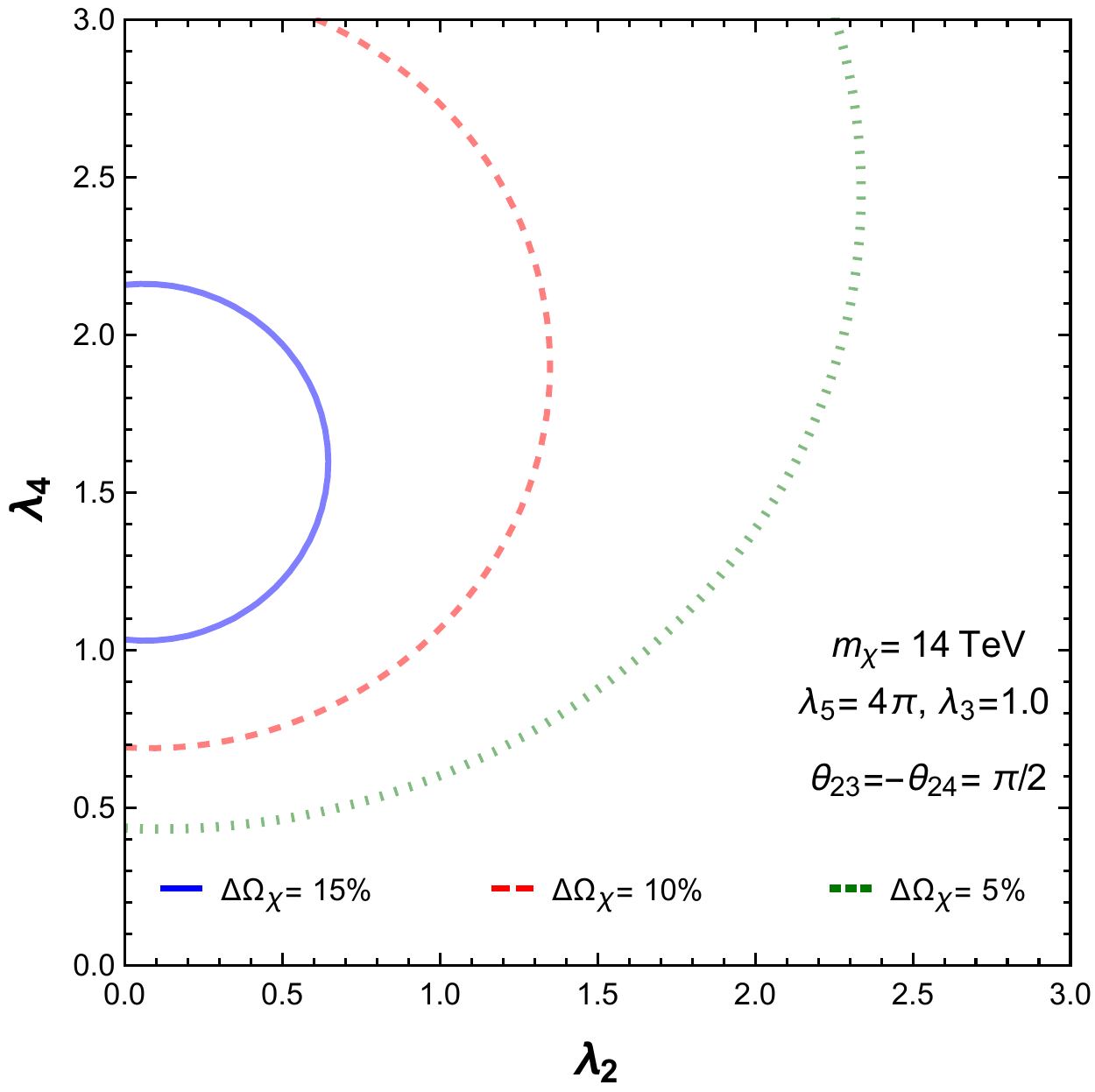}
\caption{\small{\em Role of DM self-scattering in determining the total DM relic abundance $\Omega_\chi h^2$. (Left panel:) $\Omega_\chi h^2$ as a function of the parameters $\lambda_2$ and $\lambda_4$ which determine the self-scattering rate. The maximum deviation in the relic density from the case with no self-scattering, for the choice of parameters in this figure, is $\mathcal{O}(15\%)$, where we have defined the deviation as $\Delta \Omega_\chi (\lambda_2,\lambda_4) =( \Omega_\chi (\lambda_2,\lambda_4) -  \Omega_\chi (\lambda_2=0, \lambda_4=0))/ \Omega_\chi (\lambda_2=0, \lambda_4=0)$. (Right panel:) Contours of fixed values of $\Delta \Omega_\chi$ in the $\lambda_2 - \lambda_4$ plane, with $\Delta \Omega_\chi=15\%, 10\%$ and $5\%$.}}
\label{fig:self_vs_relic}
\end{figure}

In order to further understand the impact of the self-scattering in determining the relic density, we show in Fig.~\ref{fig:self_vs_relic} (left panel) the variation in the total DM density $\Omega_\chi h^2$ as a function of the parameters determining the self-scattering rate, namely $\lambda_2$ and $\lambda_4$. For this figure, we have used a fixed ratio of these two parameters, with $2\lambda_2=\lambda_4$. All other parameters have been kept fixed. We observe that the relic abundance can vary significantly as a function of $\lambda_2$ and $\lambda_4$. For these choice of parameters, the deviation in $\Omega_\chi h^2$ from its value with no self-scattering, defined as, $\Delta \Omega_\chi (\lambda_2,\lambda_4) =( \Omega_\chi (\lambda_2,\lambda_4) -  \Omega_\chi (\lambda_2=0, \lambda_4=0))/ \Omega_\chi (\lambda_2=0, \lambda_4=0)$ takes a maximum value of around $\mathcal{O}(15\%)$. 

Such significant deviations in $\Omega_\chi h^2$ are obtained for a wide range of values for the self-scattering parameters. This is shown in Fig.~\ref{fig:self_vs_relic} (right panel), where we plot contours of fixed values of $\Delta \Omega_\chi$ in the $\lambda_2 - \lambda_4$ plane. The contours are shown for three values of $\Delta \Omega_\chi$, $15\%$ (blue solid line), $10\%$ (red dashed line) and $5\%$ (green dotted line). All other parameters are kept fixed as before. As we can see from this figure, for a fixed value of  $\Delta \Omega_\chi$, the values of $\lambda_2$ and $\lambda_4$ lie on a half-circle, and $\mathcal{O}(15\%)$ deviations are observed for a wide range of values of these parameters. Larger deviations in $\Omega_\chi h^2$ are also possible, in circles of smaller radii. Thus, these figures clearly demonstrate the significant role of the self-scatterings in determining the total relic density itself, by comparing the relic densities in the scenarios with and without DM self-scatterings. 

%%%%%%%%%%%%%%%%%%%%%%%%%%%%%%%%%%%%%%%
\section{Summary and discussion}
\label{sec:sec6}
To summarize, we studied the scenario of a complex scalar dark matter, which is a singlet under the SM gauge interactions, and is stabilized by an effective $Z_2$ reflection symmetry. We considered all the renormalizable interactions of the singlet with the SM Higgs doublet, as well as an SM singlet real scalar, including terms that preserve the reflection symmetry, but break the larger global $U(1)_\chi$ symmetry of DM number. Since the interactions of the DM with the Higgs boson are already strongly constrained, we are forced to introduce the minimal scenario with the additional real scalar which is even under the $Z_2$ symmetry $-$ in order to thermalize the DM sector with the SM sector, and thereby obtain a thermal mechanism for producing the DM density.

We find that both CP-violating and CP-conserving scattering reactions involving the DM and the real scalar field ($\phi$) can take place in the thermal bath. The CP-violating scatterings include both DM self-scattering processes as well as DM annihilations to a pair of $\phi$ particles. Since these CP-violating processes involve the $U(1)_\chi$ breaking terms, they can violate DM number as well, and thereby create a possible asymmetry in the DM sector. 

In order to understand the role of the various scattering processes in determining the DM density and composition, we set up the Boltzmann kinetic equations for the (anti-)DM yields, and study them first in terms of the thermally averaged symmetric and asymmetric scattering rates. We obtain approximate analytic solutions to the system of Boltzmann equations, which are found to be in good agreement with the exact numerical results, with a maximum difference of upto $10 \%$. We utilize the unitarity sum rules in relating the CP-violation in the three different scattering processes. These sum rules imply that the CP-violation is non-zero only in the simultaneous presence of both the self-scatterings and the pair-annihilation, and would vanish if either one of them is absent. 

We then go on to compute the relevant CP-violations, and thermally averaged reaction rates in the complex singlet scenario, including relevant Feynman graphs upto one-loop level, with the CP-violation and asymmetric reaction rates ensuing from the interference of the tree level graphs with the one-loop graphs. With these, we compute the relic abundance and asymmetry in the DM sector as a function of the coupling and mass parameters. There are broadly two different regimes observed: one in which the CP-violating pair-annihilation dominantly decides the DM relic abundance, and the other in which the CP-conserving pair-annihilation plays the major role. In the first scenario, the resulting asymmetries are found to be small, so are the correlations between the self-scattering and annihilation couplings in determining the DM density. 

In the second scenario, the asymmetries can be large, with the asymmetry parameter taking values upto $\mathcal{O}(0.5)$. Furthermore, in this scenario, we observe strong correlations between the self-scattering and annihilation processes in obtaining the required DM abundance. This role of DM self-scatterings in determining the DM density is a novel observation, which is the primary result of this paper. The reason behind this effect is that when the asymmetry in the DM sector becomes large, the asymmetric yield, which is the difference between the particle and antiparticle yields, plays a dominant role in determining the total DM density, as we also show through our analytic solutions. And this asymmetric yield is affected by both the self-scattering couplings as well as the annihilation ones that violate CP. Thus, DM self-scatterings can become important in deciding both the DM density and composition, being strongly correlated with the annihilation. This is further strengthened by the unitarity sum rules mentioned above $-$ we cannot have a non-zero asymmetry unless both these types of processes are simultaneously present. 

Although the cosmological, and the related particle physics aspects of the complex scalar singlet scenario were the focus of this paper, there are important astrophysical consequences of this scenario which should be explored in future studies. The primary direction would be to look into the indirect detection prospects in the present Universe, which requires a detailed analysis depending upon the mass of the DM $\chi$ and the real singlet scalar $\phi$. Although such an analysis is beyond the scope of this work, a few relevant comments are in order. As we have already seen, the asymmetry generated in this scenario is not maximal, and therefore, we shall have a density of both DM and anti-DM particles existing in the galaxies and clusters. The exact ratio of the densities would be dependent upon both the asymmetry parameter at the generation level, and any possible oscillation that may happen at post-freeze out epochs due to the $U(1)_\chi$ breaking mass terms, if these are present with sufficient strength. 

All the four relevant annihilation processes involving the $\chi$ and $\chi^\dagger$ particles will therefore take place, with $\chi \chi \rightarrow \phi \phi$ (and its conjugate $\chi^\dagger \chi^\dagger \rightarrow \phi \phi$) and $\chi \chi^\dagger \rightarrow \phi \phi$ leading to the production of $\phi$ particles in the final state. Since the $\phi$ particles can have a small mixing with the SM Higgs boson, they will eventually decay to SM particles through this mixing. The exact decay branching ratio to different final states will depend upon the $\phi$ mass, while the energy spectra of the decay products will depend both on the decay mode, as well as on the $\chi$ mass, thus leading to a wide range of possibilities. Given the nature of the Higgs coupling to SM particles, if kinematically allowed, the possible annihilation modes for one pair of (anti-)DM particles are then to four final state particles, drawn in two pairs primarily from the following list  $-\{\gamma \gamma$, $\tau^+ \tau^-$, $b \bar{b}$, $W^+ W^-$, $ZZ$, $t \bar{t}$ and $hh\}$. If $\phi$ is very light, for example, it may decay to $\gamma \gamma$, thereby leading to four-photon final states in (anti-)DM annihilations. 

Such a process, with $\{\chi\chi, \chi^\dagger\chi^\dagger,\chi\chi^{\dagger}\}  \rightarrow \phi \phi \rightarrow 4\gamma$ leads to an interesting feature in the spectra of the gamma ray signal. Each of the produced photons will have an energy in the band $E^{min}_\gamma \leq E_\gamma \leq E^{max}_\gamma$ and therefore the corresponding gamma-ray spectra will be of a box shape of width $\Delta E = E^{max}_\gamma-E^{min}_\gamma = \sqrt{m^2_\chi-m^2_\phi}$, see, for example~\cite{Ibarra:2012dw}. Apart from the DM mass $m_\chi$, the photon flux will depend upon the parameters $r_\infty$, $\langle \sigma v \rangle_3$ and $\langle \sigma v \rangle_A$. Therefore, compared to the standard WIMP scenario, the correspondence between the observed flux and the pair-annihilation rate is expected to be different in this model. In particular, if $\mathcal{F}_{\rm ADM}$ is the photon flux expected in our scenario from $\{\chi\chi, \chi^\dagger\chi^\dagger,\chi\chi^{\dagger}\}  \rightarrow \phi \phi$ processes and $\mathcal{F}_{\rm WIMP}$ is the corresponding flux in case of WIMPs from the same set of $\{\chi\chi, \chi^\dagger\chi^\dagger,\chi\chi^{\dagger}\}  \rightarrow \phi \phi$ processes (here, WIMPs represent the symmetric limit of the ADM in which the final particle and anti-particle number densities are the same), then we obtain the ratio between the two fluxes to be:
\begin{equation}
\frac{\mathcal{F}_{\rm ADM}}{\mathcal{F}_{\rm WIMP}}=\frac{1}{\expval{\sigma v}_A+\expval{\sigma v}_3}  \bigg[(1-r^2_\infty) \expval{\sigma v}_A + (1+r^2_\infty) \expval{\sigma v}_3 + 2 r_\infty \expval{\epsilon\sigma v}_3 \bigg]
\label{flux}
\end{equation} 
We see that while $\mathcal{F}_{\rm ADM} = \mathcal{F}_{\rm WIMP}$ for the completely symmetric case of $r_\infty=0$, $\mathcal{F}_{\rm ADM} \neq \mathcal{F}_{\rm WIMP}$ for $r_\infty \neq 0$, where the difference stems from the different number densities of the $\chi$ and $\chi^\dagger$ particles in the present epoch, while keeping their sum fixed.

Such a spectrum of photons will be discernible above the astrophysical background, for example in the Fermi-LAT satellite looking either into the galactic center or into dwarf-spheroidal galaxies~\cite{Ibarra:2012dw}. However, the relevant constraints will strongly depend upon the DM mass, and for very heavy DM particles, only a future experiment such as the CTA probing a higher photon energy range will be promising~\cite{Ibarra:2015tya}. The other possible final states do not lead to any distinct spectral feature in the gamma ray or charged particle energy distribution, and therefore would have to be searched for in the corresponding diffuse flux. As it is clear from this brief discussion, given the wide possible range of DM and $\phi$ masses, and thus the varied possible final states, there are multiple possible indirect detection probes that would require a detailed independent study. 

The complex singlet scalar constitutes one of the simplest scenarios for non-self-conjugate thermal DM. Our study shows that in its full generality, this scenario offers many novel phenomena as far as the DM cosmology is concerned. The role of DM self-scatterings, and their interplay with the other processes in driving the cosmological evolution of the DM properties is a novel effect of considerable interest which should be explored in other scenarios of thermal DM as well.

%%%%%%%%%%%%%%%%%%%%%%%%%%%%%%%%%%%%%%%

\section*{Acknowledgment}
The work of AG is partially supported by the RECAPP, Harish-Chandra Research Institute, and the work of DG is partially supported by CSIR, Government of India, under the NET JRF fellowship scheme with award file No. 09/080(1071)/2018-EMR-I, and in part by the Institute Fellowship provided by the Indian Association for the Cultivation of Science (IACS), Kolkata.

%%%%%%%%%%%%%%%%%%%%%%%%%%%%%%%%%%%%%%%%%%%%%%%

\appendix

\section{CP-violation in the $\chi+\phi \rightarrow \chi^{\dagger}+\phi$ channel}
\label{App.A}
\begin{figure}[htb!]
\includegraphics[scale=0.30]{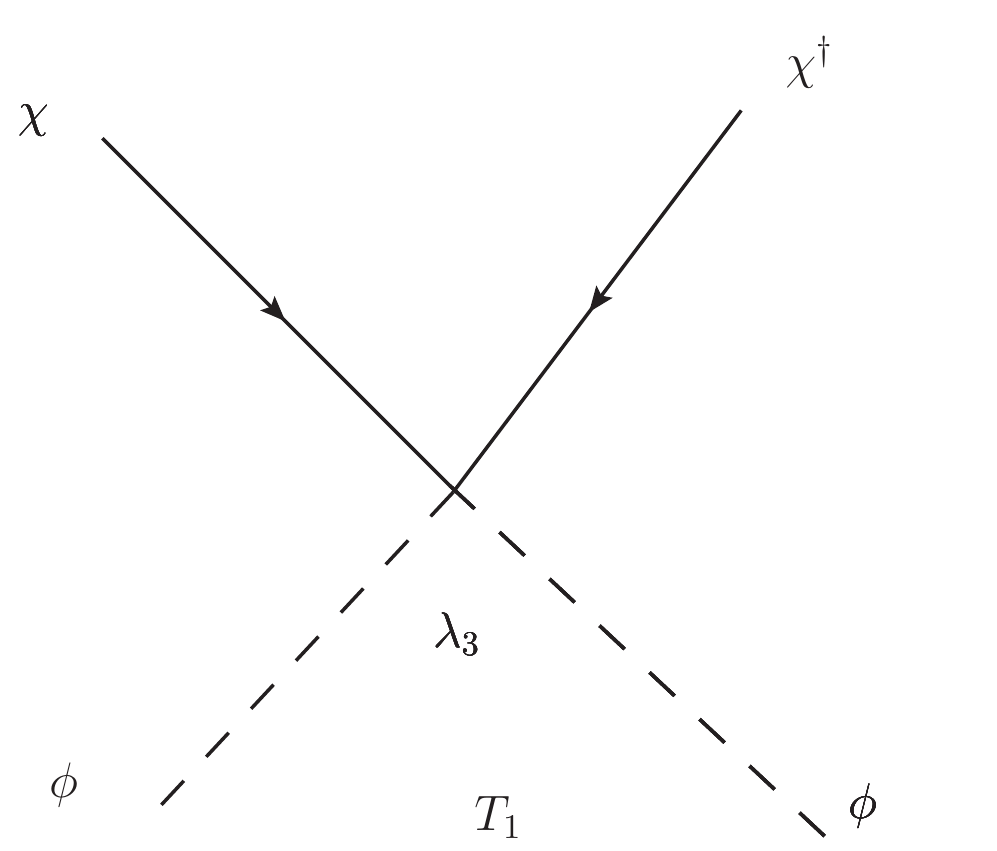}
\includegraphics[scale=0.30]{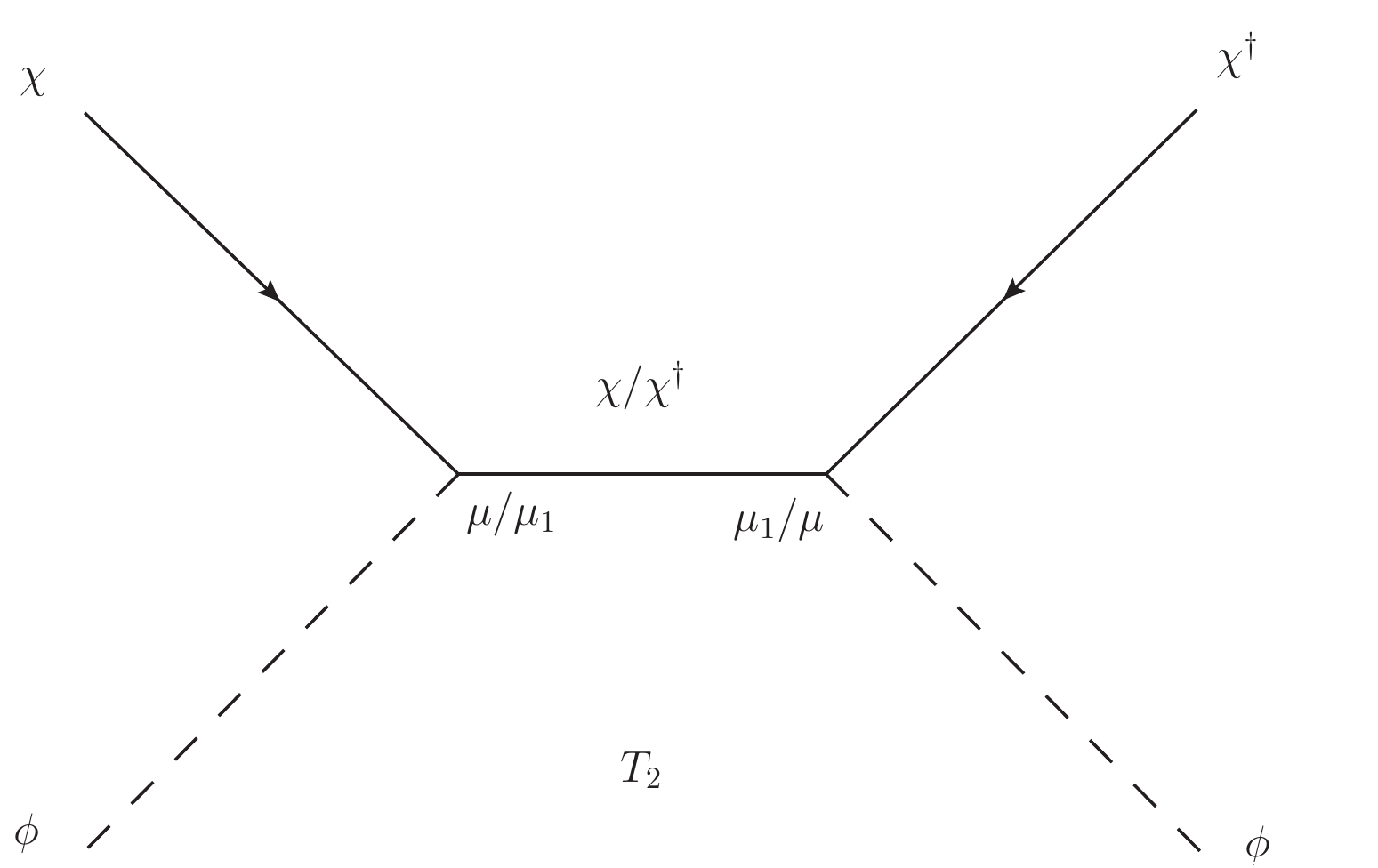}
\includegraphics[scale=0.45]{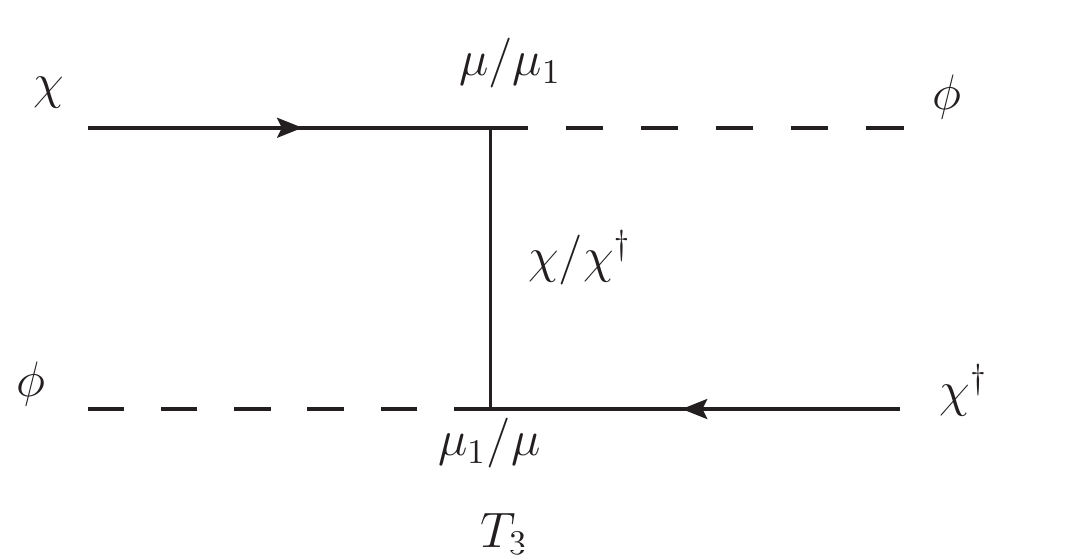}
\includegraphics[scale=0.4]{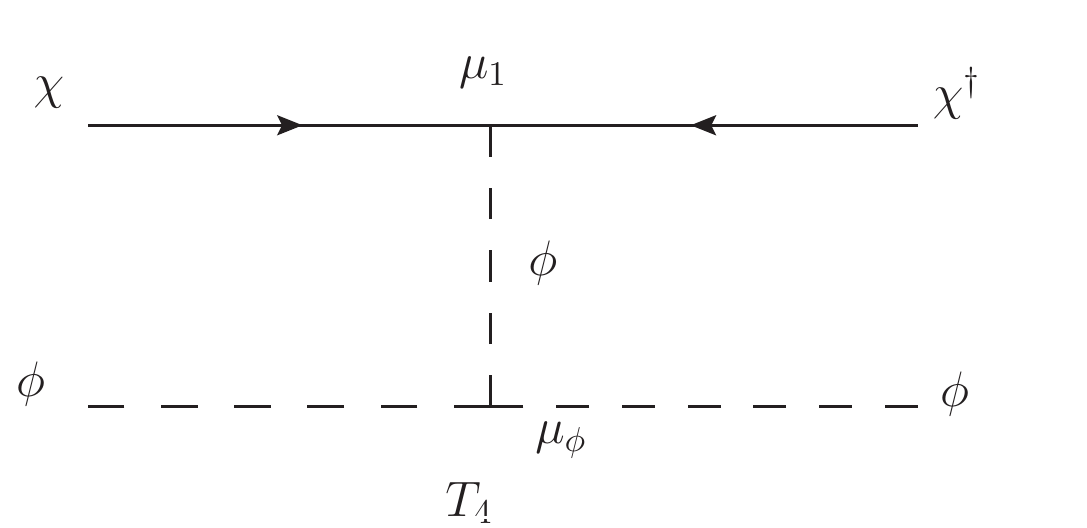}
\includegraphics[scale=0.39]{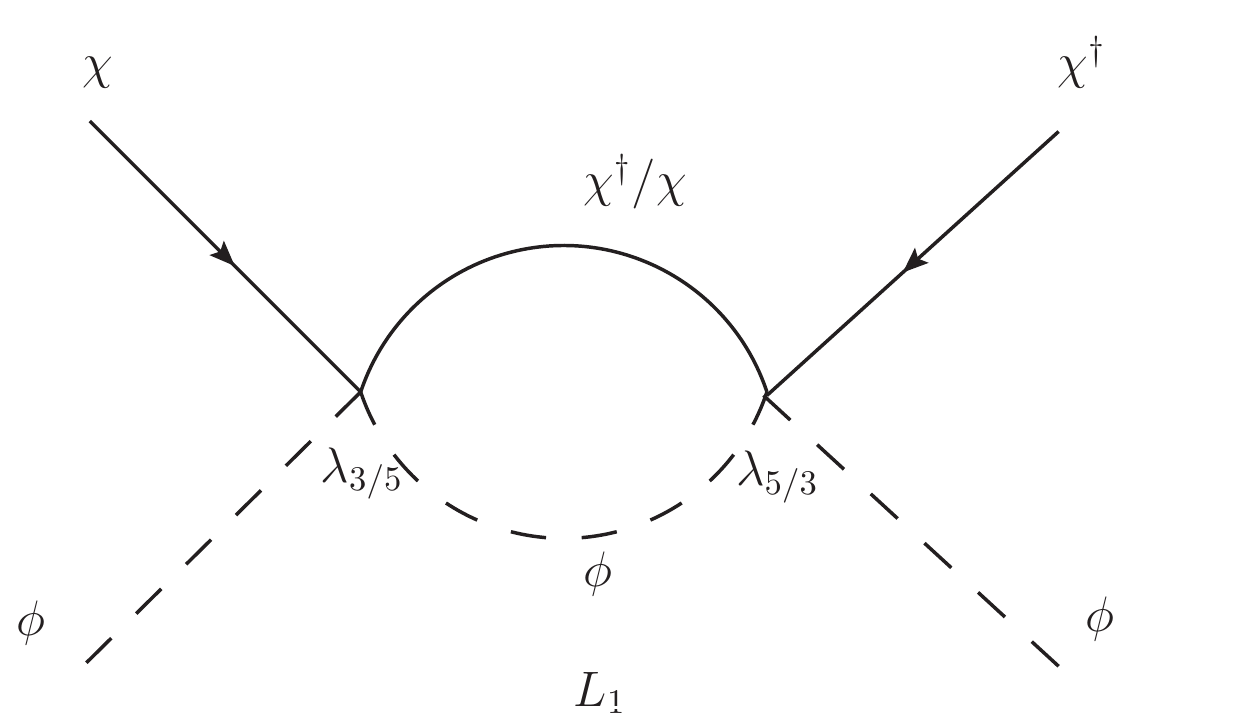}
\includegraphics[scale=0.39]{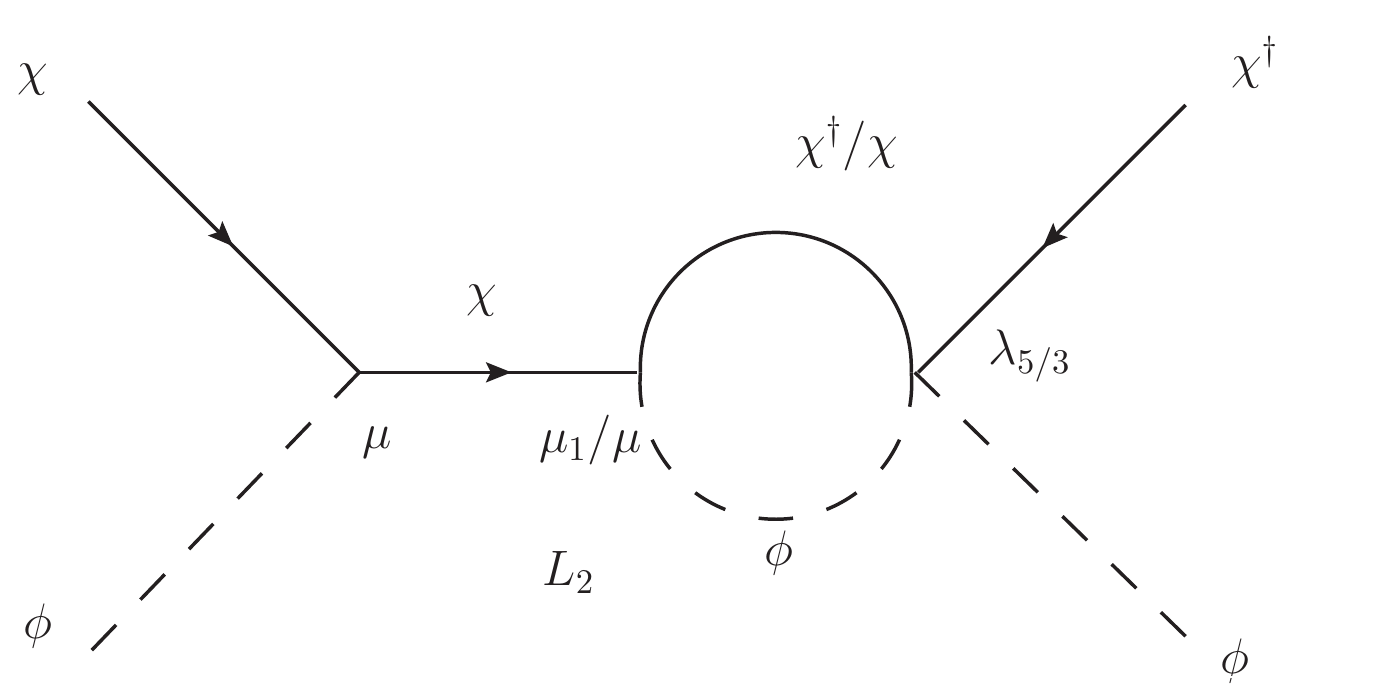}
\includegraphics[scale=0.35]{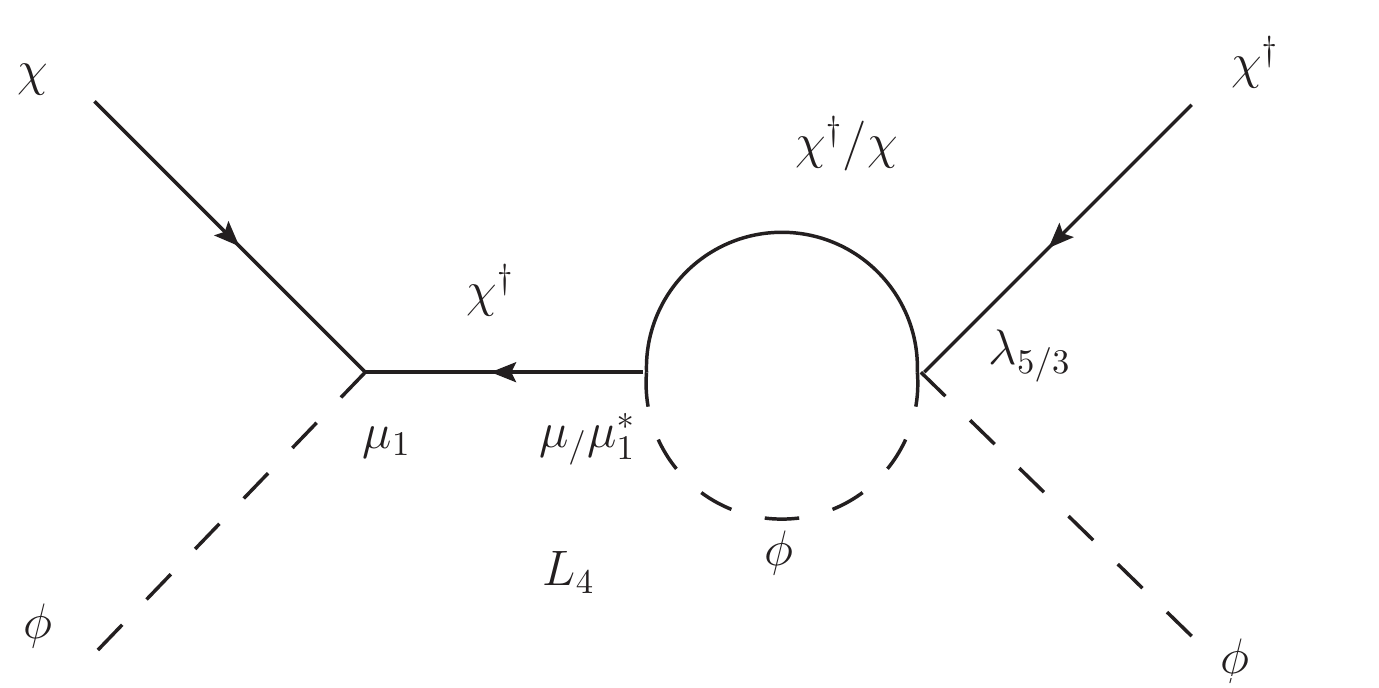}
\includegraphics[scale=0.35]{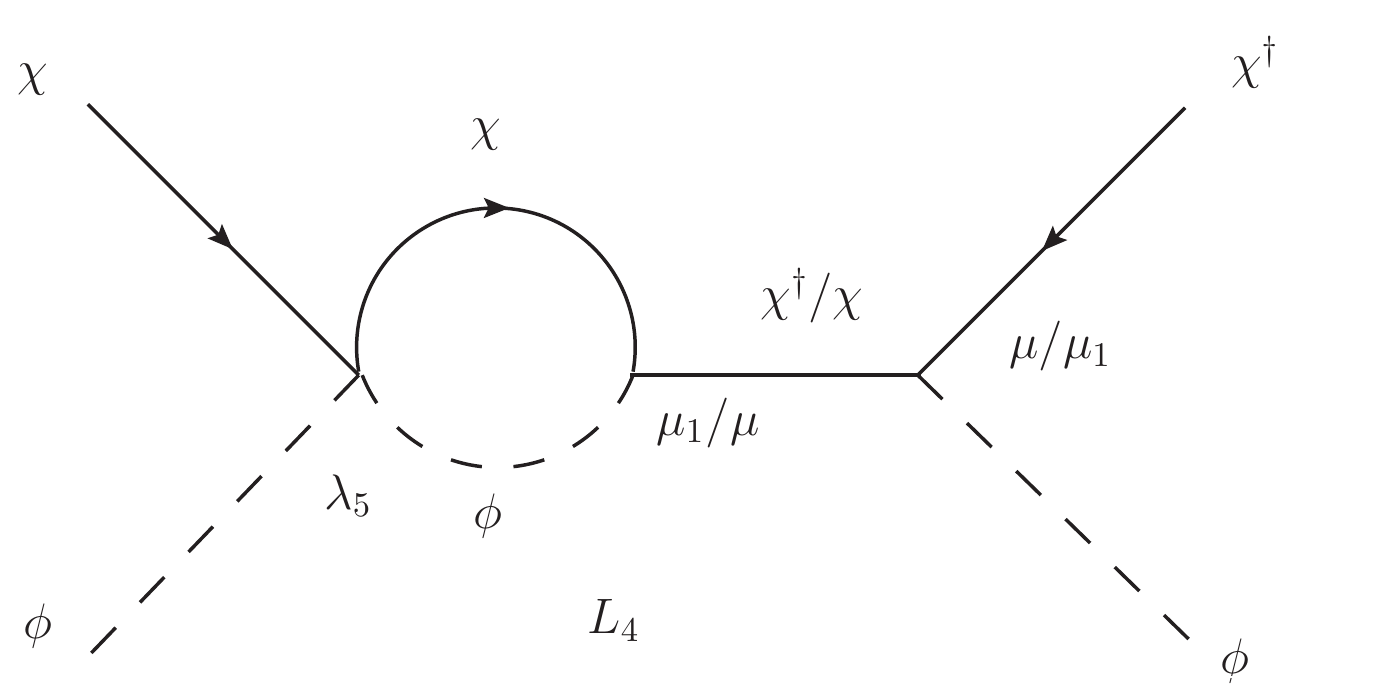}
\includegraphics[scale=0.35]{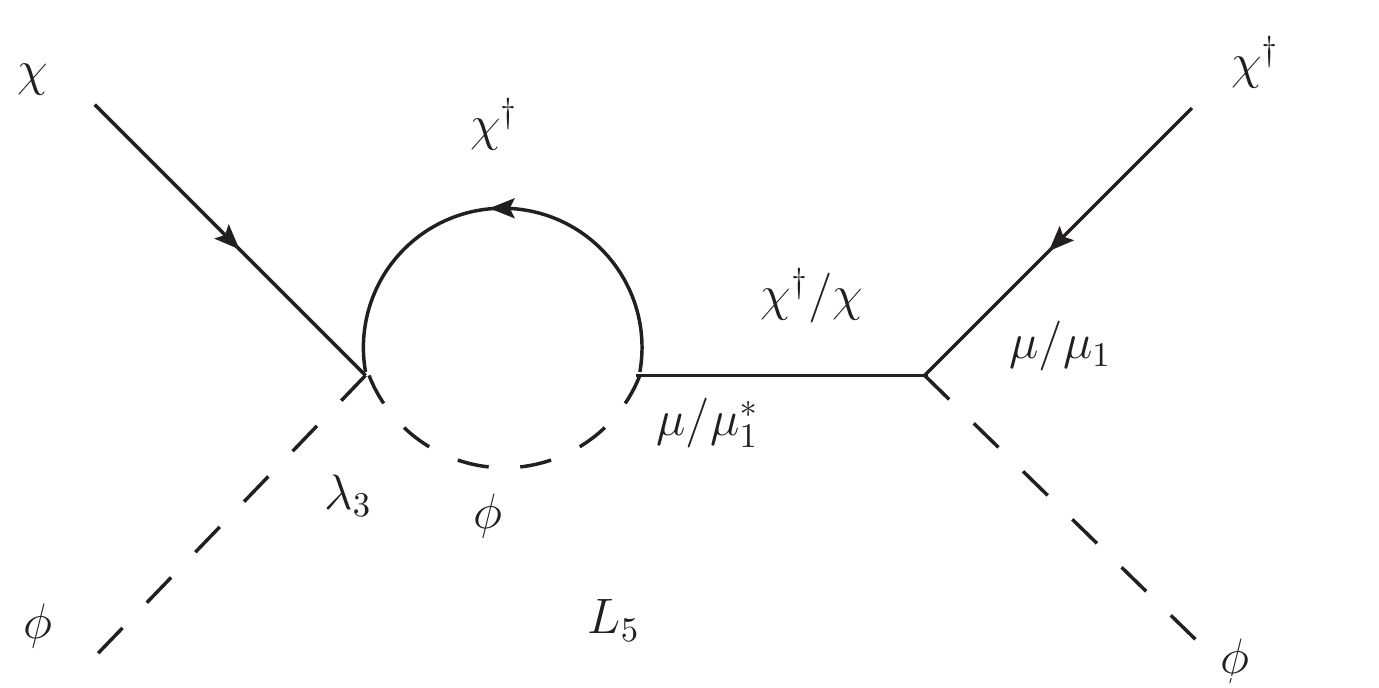}
\includegraphics[scale=0.36]{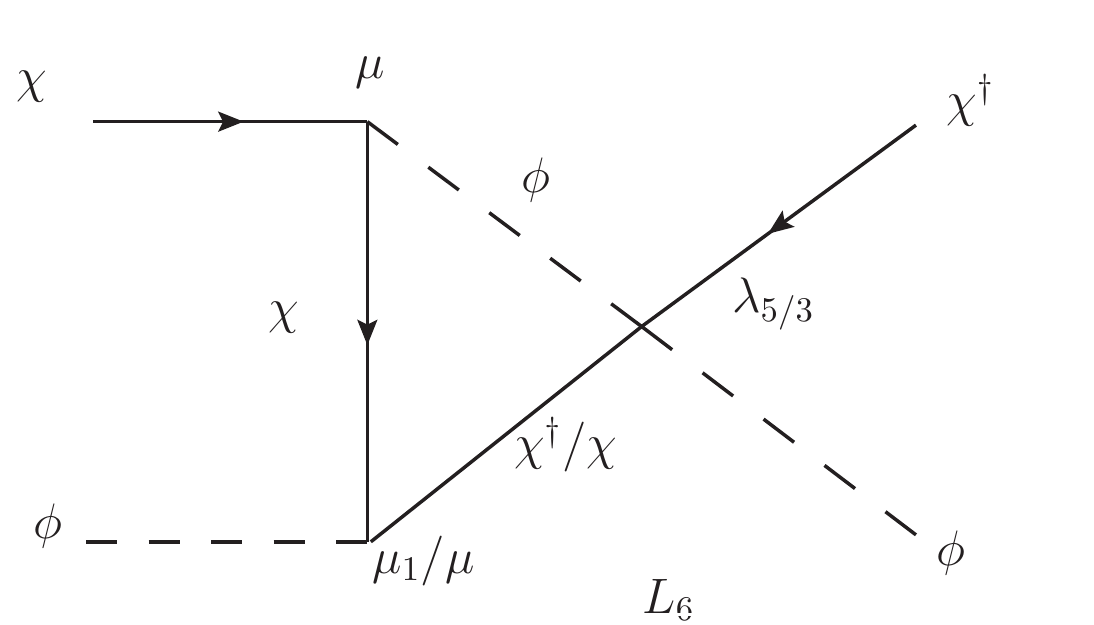}
\includegraphics[scale=0.36]{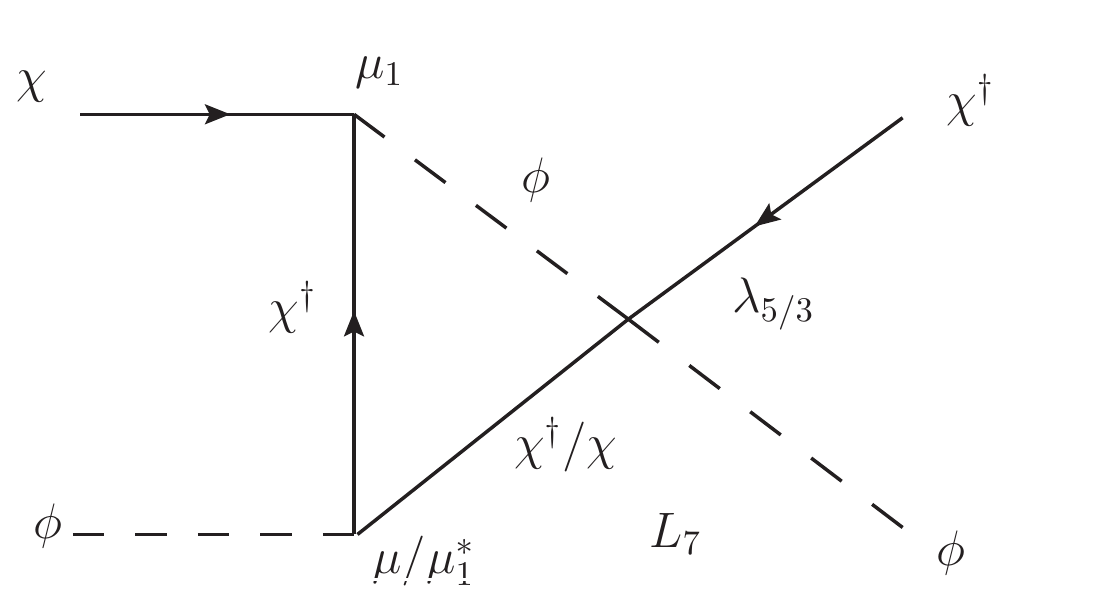}
\includegraphics[scale=0.36]{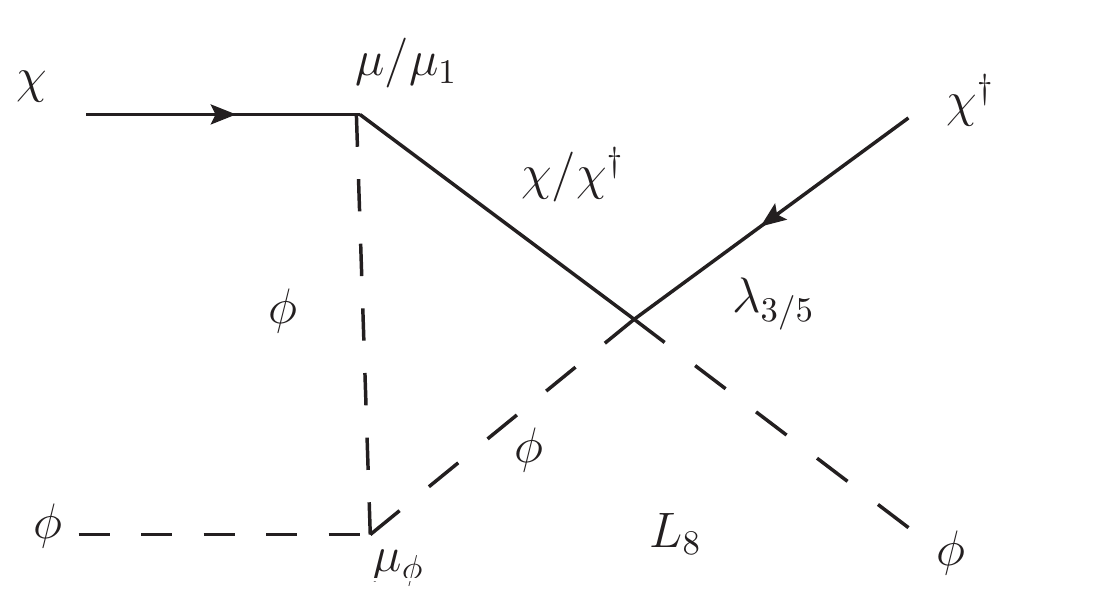}
\includegraphics[scale=0.36]{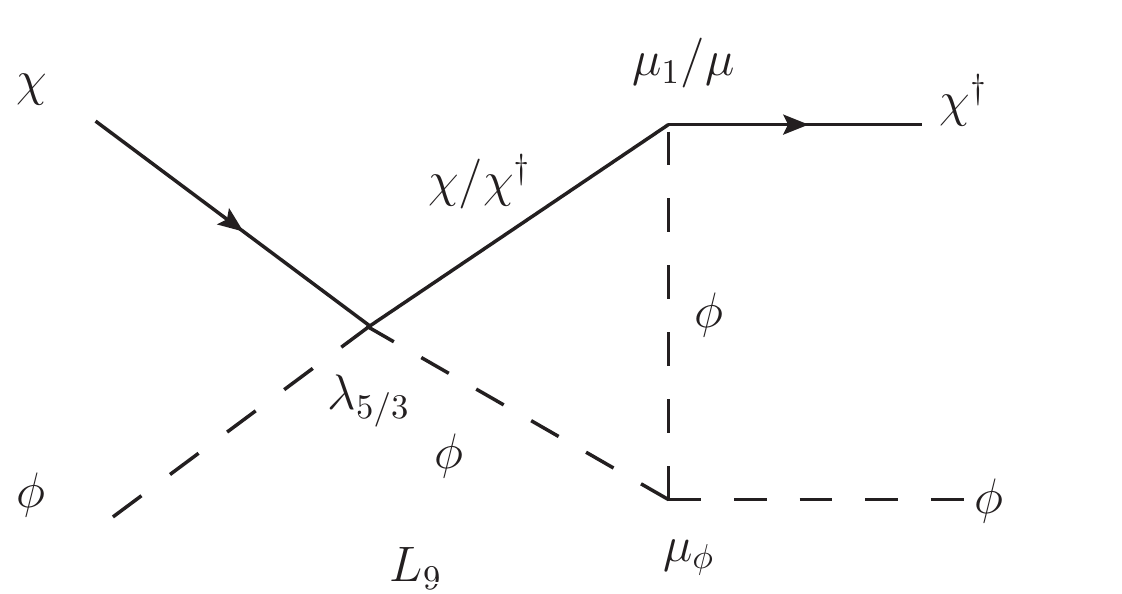}
\includegraphics[scale=0.36]{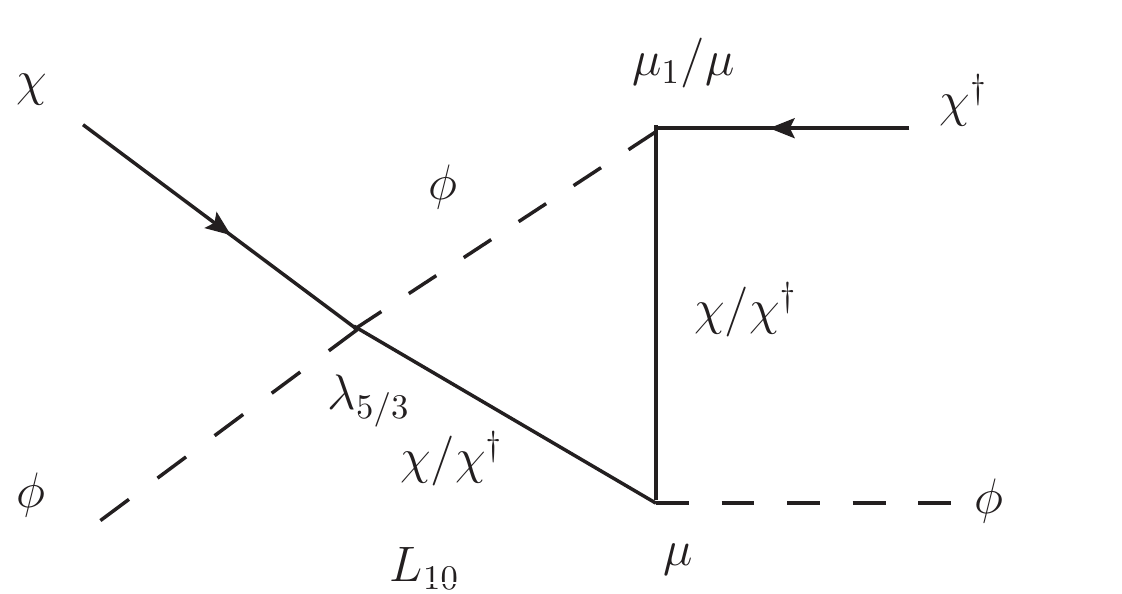}
\includegraphics[scale=0.36]{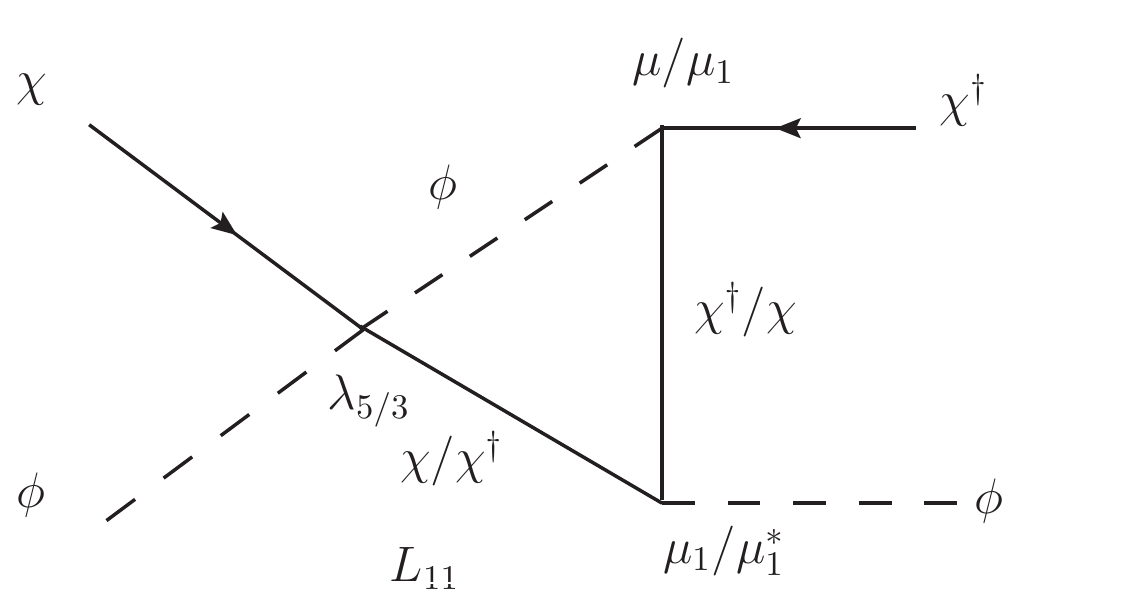}
\caption{\small{\em Relevant tree level and one-loop Feynman diagrams for the CP-violating DM-anti DM conversion process, $\chi+\phi \rightarrow \chi^{\dagger} + \phi$. Apart from $T_1$ and $T_4$, each diagram represents two distinct topologies, depending upon whether the intermediate state  is a particle or an antiparticle state. The particle-flow arrows and the relevant couplings are chosen accordingly.}}
\label{fig:diag10}
\end{figure}

In this appendix, we describe the computational details of the CP-violation in the process $\chi+\phi \rightarrow \chi^{\dagger}+\phi$, the results of which were quoted in Secs.~\ref{sec:sec2} and ~\ref{sec:sec4}. The relevant Feynman diagrams for the process $\chi+\phi \rightarrow \chi^{\dagger}+\phi$ are shown in Fig.~\ref{fig:diag10}. The CP-violation, resulting  from the interference of the tree-level diagrams with the one-loop ones, can be divided into two categories. The first category is the interference of the contact interaction-induced tree diagram $T_1$ with the one-loop ones, which gives
\begin{align}
\int dPS_2\left(|M|^2_{\chi\phi\rightarrow \chi^\dagger\phi}-|M|^2_{\chi^{\dagger}\phi \rightarrow \chi\phi}\right)_{T_1-{\rm Loop}}= 4\, \rm Im(\lambda^*_3\mu_1)\int dPS_2\,\bigg[\mu\lambda_5\, Im \big(L_2+L_3+2\,L_4 \nonumber\\
+L_6+L_7+L_{10}+L_{11}\big)+\mu_\phi \lambda_5\, \rm Im \left(L_8+L_{9}\right) \bigg].
\end{align}
Using the Cutkosky rules \cite{Peskin:1995ev}, we find that $\rm Im L_2=Im L_3=Im L_4$, $\rm Im L_6=Im L_7=Im L_{10}=Im L_{11}$ and  $\rm Im L_8=Im L_9$. Thus, the above expression can be simplified as 
\begin{align}
\int dPS_2\left(|M|^2_{\chi\phi\rightarrow \chi^\dagger\phi}-|M|^2_{\chi^{\dagger}\phi \rightarrow \chi\phi}\right)_{T_1-{\rm Loop}}= 4\, \rm Im(\lambda^*_3\mu_1)\int dPS_2\,\bigg[4\,\mu\lambda_5\, Im \left( L_2+L_6\right) \nonumber \\
+2\,\mu_\phi \lambda_5\, \rm Im \left(L_8\right) \bigg]  
\label{eq:eq.A3} 
\end{align}
Similarly, the CP-violation coming from the interference of the other three tree-level diagrams $T_2,T_3$ and $T_4$ with the loop-amplitudes is given by
\begin{align}
\sum^4_{i=2}\int dPS_2\left(|M|^2_{\chi\phi\rightarrow \chi^\dagger\phi}-|M|^2_{\chi^{\dagger}\phi \rightarrow \chi\phi}\right)_{T_i-{\rm Loop}}= - 4\, \rm Im(\lambda^*_3\mu_1)\int dPS_2\,\bigg[4 \mu \lambda _5 \rm Im\, L_1\,\bigg(\frac{1}{s-m_\chi^2} \nonumber\\
+\frac{1}{t-m_\chi^2}\bigg)+2 \mu_\phi \lambda_5\, \rm Im\, L_1\frac{1}{t-m^2_\phi}+ \mathcal{O}(|\hat\mu/m_\chi|^3)\bigg].
\label{eq:eq.A4} 
\end{align}
From the explicit calculations of the integrals in Eq.\ref{eq:eq.A3} and Eq.\ref{eq:eq.A4}  we find the following relation 
\begin{align}
\int dPS_2\,\bigg[4\,\mu\lambda_5\,\rm Im \left( L_2+L_6\right)+
2\,\mu_\phi \lambda_5\, \rm Im \left(L_8\right) \bigg]=\int dPS_2\,\bigg[4 \mu \lambda _5 \rm Im\, L_1\,\left(\frac{1}{s-m_\chi^2} +\frac{1}{t-m_\chi^2}\right )\nonumber\\
+2 \mu_\phi \lambda_5\, \rm Im\, L_1\frac{1}{t-m^2_\phi}\bigg],
\end{align}
where, $s$,$t$ are the Mandelstam variables. This shows that the leading terms cancel identically between the contribution from $T_1$ and the combined contributions from $T_2,T_3$ and $T_4$, whereas the surviving sub-leading terms are found to be 
\begin{align}
\sum^4_{i=1}\int dPS_2\left(|M|^2_{\chi\phi\rightarrow \chi^\dagger\phi}-|M|^2_{\chi^{\dagger}\phi \rightarrow \chi\phi}\right)_{T_i-{\rm Loop}} \propto{\rm Im(\lambda^*_3\mu_1)} \times {\rm ~Terms ~of}~\mathcal{O}({\hat\mu^3/m_\chi^4}),
\end{align}
where, $\hat{\mu}$ is either $\mu$, $\mu_1$ or $\mu_\phi$.

%%%%%%%%%%%%%%%%%%%%%%%%%%%%%%%%%%%%%%%%%

\end{document}